# Open Source Software Sustainability Models: Initial White Paper from the Informatics Technology for Cancer Research Sustainability and Industry Partnership Work Group

Ye Y (U Pitt), Boyce RD (U Pitt), Davis MK (U Pitt), Elliston K (tranSMART Foundation & Axiomedix, Inc.), Davatzikos C (U Penn), Fedorov A (Harvard), Fillion-Robin JC (Kitware Inc.), Foster I (U Chicago), Gilbertson J (U Pitt), Heiskanen M (NCI ITCR), Klemm J (NCI ITCR), Lasso A (Queen's University), Miller JV (GE Research), Morgan M (RPCI), Pieper S (Isomics Inc.), Raumann B (U Chicago), Sarachan B (GE Research), Savova G (Harvard), Silverstein JC (U Pitt), Taylor D (U Pitt), Zelnis J (U Pitt), Zhang GQ (UT Health) and Becich MJ (U Pitt).

**Abstract**: The Sustainability and Industry Partnership Work Group (SIP-WG) is a part of the National Cancer Institute Informatics Technology for Cancer Research (ITCR) program. The charter of the SIP-WG is to investigate options of long-term sustainability of open source software (OSS) developed by the ITCR, in part by developing a collection of business model archetypes that can serve as sustainability plans for ITCR OSS development initiatives. The workgroup assembled models from the ITCR program, from other studies, and via engagement of its extensive network of relationships with other organizations (e.g., Chan Zuckerberg Initiative, Open Source Initiative and Software Sustainability Institute). This article reviews existing sustainability models and describes ten OSS use cases disseminated by the SIP-WG and others, and highlights five essential attributes (alignment with unmet scientific needs, dedicated development team, vibrant user community, feasible licensing model, and sustainable financial model) to assist academic software developers in achieving best practice in software sustainability.

**INTRODUCTION**

The Informatics Technology for Cancer Research (ITCR) program[1] is a program of the National Cancer Institute (NCI) established in 2012 to create an ecosystem of open source software (OSS) that serves the needs of cancer research. ITCR is an NCI program that supports informatics technology development initiated by cancer research investigators and includes all four NCI extramural divisions: Cancer Biology, Cancer Control and Population Science, Cancer Prevention, and Cancer Treatment and Diagnosis. The coordinating body for ITCR is the NCI Center for Biomedical Informatics and Informatics Technology.

The specific goals of ITCR include: 1) promote integration of informatics technology development with hypothesis-driven cancer research and translational/clinical investigations; 2) provide flexible, scalable, and sustainable support using multiple mechanisms matched to various needs and different stages of informatics technology development throughout the development lifecycle; 3) promote interdisciplinary collaboration and public-private partnership in technology development and distribution; 4) promote data sharing and development of informatics tools to enable data sharing; 5) promote technology dissemination and software reuse; 6) promote communication and interaction among development teams; and 7) leverage NCI program expertise and resources across the institute and bridge gaps of existing NCI grant portfolios in informatics.

The scope of the ITCR program is to serve informatics needs that span the cancer research continuum. The ITCR program provides a series of funding mechanisms that support for informatics resources across the development lifecycle, including the development of innovative methods and algorithms (R21), early stage software development (R21), advanced stage software development (U24), and sustainment of high-value resources (U24) on which the cancer research and translational informatics community has come to depend. The current funding opportunities are available from the ITCR website.[2]

This series of funding mechanisms is innovative and unique in all the NIH institutes and Centers. It addresses a fundamental need to create computational infrastructure that is interoperable and collaborative, linking many of the informatics and computational biology teams doing translational informatics. The ITCR ecosystem has grown substantially and now includes 66 funded efforts that are highly collaborative, as evidenced by its "connectivity map" (Figure 1).

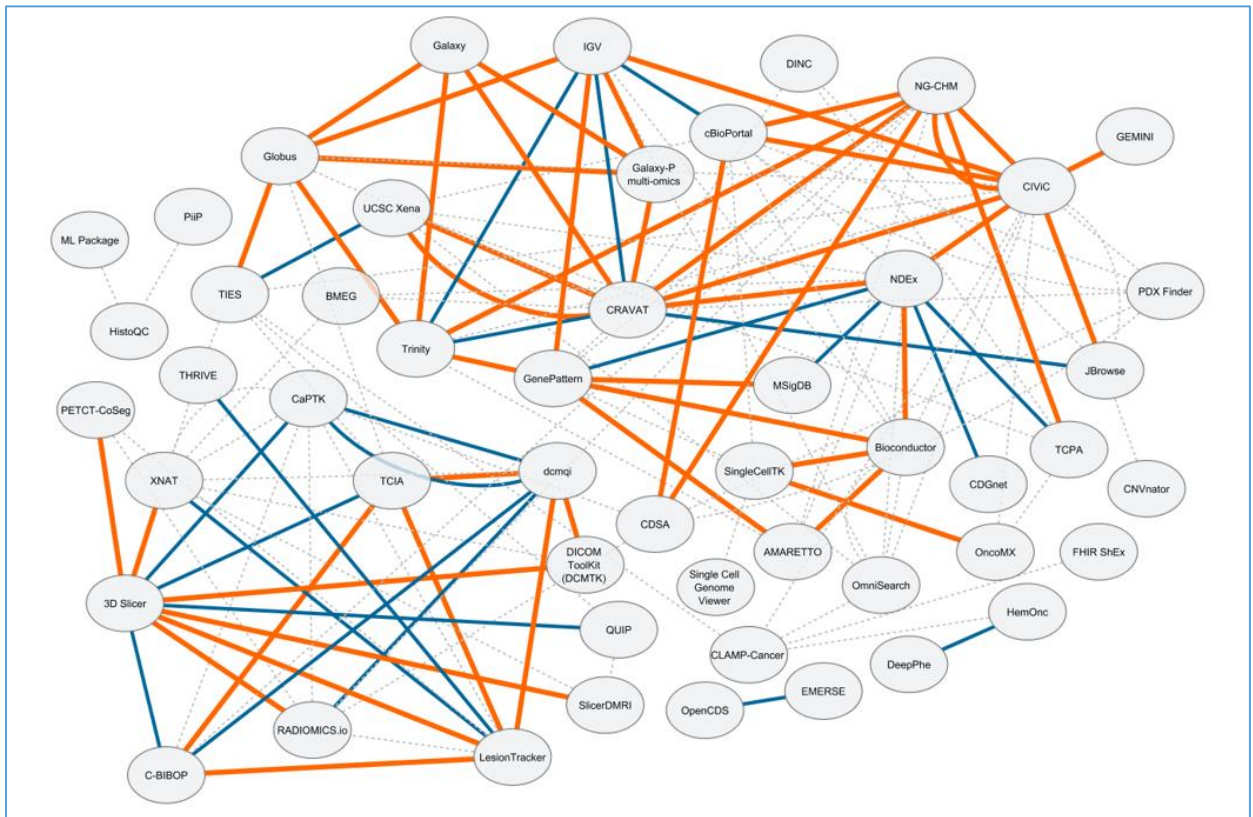

*Figure 1. The map of ITCR projects. This map is copied from the NDEx[3] website.[4] In this map, each node represents a project funded under the ITCR. Edges among these nodes represent connections among projects: existing connections (orange solid lines), ongoing connections (blue solid lines), and proposed connections (grey dashed lines).*

      Several informatics efforts across the NIH have emphasized creating an approach to open source software sustainment. The programs include the informatics efforts linking the National Center for Advancing Translational Sciences, Clinical and Translational Science Awards (CTSA) with programs from the NIH Office of the Director, including the Big Data to Knowledge (BD2K)[5] and the Data Science[6] program, and most recently, the All of US Precision Medicine Initiative.[7] Phil Bourne and colleagues[8] outlined a set of goals that describe the challenges of sustainable OSS (e.g., mostly developed as prototype software, not scale to terabytes of data, and lack of scientific attribution for software development), which in part the ITCR is nicely positioned to address for the cancer research community.

      As the ITCR program moves into its second phase, it faces the challenge of long-term sustainability of software being developed by its grantees. Whether viewed from the angle of a single funded project, or across all ITCR-funded groups, some of the software will naturally

"graduate" upon reaching maturity, in order to leave room for continuing innovation by the program. As mature projects often lead to complex and successful products based on years of investment of human effort, funding, and cumulative expertise, they need to move into their next phase of support, rather than risking being abandoned.

Addressing this challenge was the primary task of the ITCR Sustainability and Industry Partnership Working Group[9] (ITCR SIP-WG), which was convened in 2019. The working group initially set the goal of addressing four major topics of interest to the translational cancer informatics community: 1) publish a collection of case studies of successfully disseminated software products supported by open source licenses, to provide practical examples of approaches proven viable for licensing and sustainability; 2) develop a workflow/decision tree to support informed decision making, consistent with ITCR expectations and the future licensing needs of open source tools; 3) provide a licensing consultancy service, in collaboration with the ITCR program; and 4) develop a collection of business model archetypes which can serve as starting templates, to formally document dissemination and sustainability plans for new software development initiatives. The ITCR licensing resources will represent "best practice" approaches and will leverage our extensive network of relationships with organizations such as the Open Source Initiative, the Software Sustainability Institute, and the Chan Zuckerberg Initiative to maintain relevant knowledge in this field. The first major topic—publication of case studies—listed above is the subject of this manuscript. The remaining three topics will be the focus of future whitepapers and manuscripts by the ITCR SIP-WG.

**LITERATURE REVIEW**

Several papers discussed sustainability models.[10-12] Aartsen et al.[12] described two models for sustaining digital assets from public-private partnerships in medical research. The ***not-for-profit organization model*** uses, for example, a foundation (also discussed by Kuchinke et al.[10]), as the backbone organization to assure the maximum value of the assets. The Apache Software Foundation is one such example. An advantage of non-profits is that they can take a long-term view. The sustainability of the non-profits can be mitigated through memberships. The concept of a foundation has the advantage that the development of the artifact is strongly influenced by academic users, so its design can be focused on academic goals instead of commercial ones. The disadvantage of the not-for-profit organization model is its dependency on one organization for all digital assets. The ***distributed network model*** is built on the premise that the individual partners who contributed to the development of the digital assets have a stake in seeing them sustained and gaining future value through further development. The disadvantage of that model is the conflicting missions of research and industry – organizations with a research mission do not focus on producing digital assets ready to be commercialized by industry.

Gabella et al.[13] added several other models for the sustainability of assets, including four non-commercial models. The ***national funding model*** funds the infrastructure directly through non-cyclical funding programs. In the ***infrastructure model***, funding agencies set aside a fixed percentage of their research grant volumes to be redistributed to core data resources according to well-defined selection criteria. In the ***institutional support model***, funds are provided from their own organization to sustain the data assets. The ***donation model*** depends on philanthropic funding. Gabella et al.[13] also mentioned six commercial models. The ***content licensing/industrial support model*** requires commercial users to pay a fee for access and for-profit use, whereas the assets are free for non-commercial users (also discussed in Kuchinke et al.[10]). The ***user subscription model*** (also discussed in Chang et al.[11]) relies on a subscription for a set period of time. The ***freemium model*** (also discussed in Chang et al.[11]) provides a core that is free, with add-ons for a fee. The ***razor and blades model*** offers a free initial trial ("razor") that encourages continuing future purchases of follow-up services ("blades") (also discussed in a Wikipedia introduction of Business models for open source software[14] as one of the commercial models). The ***mixed models*** rely on diversified multiple funding streams. A common mixed model practice is the combination of OSS projects supported by services provided by companies that help customers with installation, configuration and troubleshooting; Linux is a familiar example. The Linux model, however, relies on a large user base which might not be necessarily the case with biomedical research tools.

In addition, the ***Macro R&D infrastructure*** discussed in Chang et al.[11] is based on funding that comes from governmental research grants or from research grants from local or international partner institutes. The ***split licensing model***[14] offers the free version under a GPL license and a commercial version with its own license that does not allow software redistribution (e.g., MySQL, openClinica).

Many interesting papers discussed the importance of the strength and the health of the community behind a software product.[15-17] Laffaldano et al.[15] used the sleep stage metaphor to describe the developer cycles – the wake stage is when developers are active in the project, the sleep stage when developers pause their package commit activity, and the dead stage when developers abandon the project. The authors further explored the reasons for the stage transitions and identified personal factors (e.g., life event, financial, change of interest) and project factors (e.g., social, changes in the project, role change). Atiq et al.[16] suggested sponsoring of open source projects in various ways as more and more proprietary firms participated, sponsored, and offered their developers for open source projects. Jimenez et al.[17] provided four recommendations for a sustainable open source project: (1) make source code publicly accessible from day one; (2) make software easy to discover by providing software metadata via a popular community registry; (3) adopt a licensing system that complies with the licenses of third-party dependencies; and (4) define clear and transparent contribution, governance, and communication processes.

Nyman and Lindman[18,19] discussed code forking (the implementation of an existing code base in a separate project), governance, and sustainability of open source software. The right to fork code is built into the very definition of open source. They argued that forking code can revive community interest in a project or provide an alternative to acquisitions which was the case with MySQL after Oracle's acquisition of Sun Microsystems. The MySQL code was forked under a different name, MariaDB, due to concerns regarding the governance and future openness of the MySQL code. Nyman and Lindman stated that "given that forking ensures that any project can continue as long as there is sufficient community interest, we have previously described forking as the 'invisible hand of sustainability' in open source."

**METHOD**

To select a collection of case examples of successfully disseminated software products, we, ITCR SIP-WG, conducted a survey among the members of the working group. We asked each member to provide the "best three" examples of sustainable OSS to serve as role models for ITCR open source projects. The survey was completed by 13 participants, most of whom are authors of this whitepaper. Twenty-two OSS use cases were provided by this ITCR SIP-WG vote and the top ten voted tools were then assigned to authors to profile models of success in sustainability (Table 1).

**Table 1 – Open Source Software Models Profiled**

| OSS Use Case | Link |
|---|---|
| 3D Slicer | https://www.slicer.org/ |
| Bioconductor | https://www.bioconductor.org/ |
| Cytoscape | https://cytoscape.org |
| Globus | https://www.globus.org/ |
| i2b2 tranSMART | https://transmartfoundation.org/ |
| ITK | https://itk.org |
| Linux | ttps://www.linuxfoundation.org/projects/linux/ |
| OHDSI | https://www.ohdsi.org/ |
| R | https://www.r-project.org/ |
| REDCap* | https://www.project-redcap.org/ |

*REDCap provides non-profit end-user license agreement, but its code base is not open to individual developers.*

After literature reviews and discussions in the ITCR-WG, we determined that each OSS use case should be profiled according to recommendations by Nesbit[20] in "What does a sustainable open source project look like?". Each of the top ten OSS use cases was profiled in terms of sustainability aspects that include governance, documentation, code quality, support, ecosystem collaboration, security, legal, finance, marketing, and dependency hygiene.

## RESULTS

In this section, we use the ten OSS use cases as examples in the ten sustainability aspects. Full descriptions of these use cases are available in a supplementary file.

**Governance**

Most of the ten OSS use cases have a management committee and a technology development team. ITK and REDCap have established consortiums. Three models (i2b2 tranSMART, R, and Linux) have established foundations. Stakeholders usually choose a consortium management model during the early stages of software development. In a consortium model, members have stronger control over the direction of development. A consortium management model may later migrate to a foundation model. In a foundation model, the organization considers the interests of all stakeholders, encouraging more new contributors and users to participate in the software development testing process.

A few OSS tools have provided governance guidelines publicly. 3D Slicer has listed contribution guidelines and a roadmap for development. The i2b2 tranSMART foundation has published bylaws online.

Most use cases have yearly regular meetings among stakeholders to make operational decisions and set development priorities. For example, 3D Slicer's core developers and users meet in person twice a year, and Globus has an annual conference for its users and subscribers. Subcommunities usually have more frequent regular meetings. For example, the technical advisory board of Bioconductor meets monthly to develop strategies to ensure the long-term technical suitability of core infrastructure. To reach a broader group of potential developers and users, some models (3D Slicer, i2b2 tranSMART) provide completely open communication channels, such as online forums and recorded webinars.

From public information on these ten OSS use cases, we do not know the exact size of each core development team or the individual assignments on core infrastructure. If there is a single person handling the complicated details of a critical component, an OSS project will be adrift quickly after losing him/her.

**Documentation**

All ten OSS use cases provide documentation to users in various formats, such as user guidebooks (ITK, Linux, and R), Wiki pages (3D Slicer, i2b2), tutorials (Bioconductor, Globus, REDCap, Cystoscape, tranSMART, 3D Slicer, and OHDSI), and YouTube videos (3D Slicer, Cystoscape).

Many OSS use cases further provide documentation to new developers to encourage new contributions to OSS extensions. For example, Bioconductor offers three levels of documentation—workflows, package vignettes, and function manual pages—encouraging users to become developers, making their own algorithms and approaches available to others. Similarly, the Cystoscape "App Ladder" teaches essential skills in developing apps. R provides various fully developed documentation, adequately covering two types of development: writing R extensions and developing R itself (by providing internal structure and coding standards).

**Code quality**

A few OSS use cases go through rigorous tests before public releases. Before propagating latest packages to user-facing repositories, Bioconductor developers always conduct tests to ensure overall package integrity and integration with current versions of package dependencies. 3D Slicer has established infrastructure to continuously run about 700 tests for the core application. The testing results are available publicly. The quality control of some of extensions of 3D Slicer is a little weaker than that of the core application. Extension contributors themselves manage code quality and tests, and 3D Slicer's core developer team does not enforce or verify these extensions. Cytoscape developers use Jenkins to build software projects continuously, and test packages thoroughly before releasing them. Globus employs a continuous integration environment, automated tests, multiple pre-release environments, and documented, standardized, human quality assurance testing to ensure code quality, with at least one engineer other than the code author reviewing code before releasing to production. A few models (3D Slicer, ITK, and R) enforce a consistent coding style.

**Support**

Most OSS use cases provide various supports to users and new developers. For example, the OHDSI community provides two support channels. The community-based discourse forum provides support for implementing OHDSI tools, proposing or participating in network research studies, and requesting information on OHDSI related topics. The GitHub project sites of OHDSI manage specific technical questions through tickets that anyone can issue. Globus has several support options: online self-help tools, listserv groups, and a ticketing submission system with a responsive support team. R mainly relies on online self-help tools, FAQ listings, and subscription-based email lists, including a general R-help email list, an R-developer list, and an R-package-developer list. While these models provide various supporting channels, Linux and Cytoscape mainly rely on dedicated channels (Linux: LF JIRA; Cytoscape: a specific Help Desk).

Not all supports are free. For example, ITK has three-way supports: (1) ITK's discourse forum enables discussion and mutual help among users, and dedicated volunteers usually provide detailed example code; (2) the NIH has continued to provide maintenance contracts for bug fixes, incremental improvements, and a moderate level of user support (maintenance has typically been performed by Kitware, providing continuity and expertise); (3) Kitware also offers commercial

ITK support for a fee. Globus operates for free support lists and a ticketing system, but guarantees subscribers a one-business-day response time on support tickets.

Surprisingly, free support is often available in a timely manner. One good example is 3D Slicer, which had over 13,000 forum posts in 2018, with average response time of less than two days (or less than eight hours during weekdays). Notably, in the case of 3D Slicer, support may be provided either by core developers or by the experienced members of the user community. Public forums can be extremely active. For example, Bioconductor has over 100 visitors per hour.

**Ecosystem collaboration**

Ecosystem collaborations are usually organized by workgroups, conferences, networks, and vibrant community forums. Most of the ten OSS use cases have complicated dependencies, but the robustness of dependencies is not always clear.

Danny Crichton[21] has pointed out the potential danger of complicated dependencies: "Blackbox can make it difficult to see that there are far fewer maintainers working behind the scenes at each of these open source projects than what one might expect." Thus, it is critical to provide transparent information about the dependency tree of libraries. 3D Slicer is a good example. It publicly provides an extensive list of dependencies.

**Security**

Linux has strong security features. The Linux kernel allows administrators to improve security at the lowest level by modifying attributes of the kernel's operation, building additional security measures into the kernel to avoid common buffer overflow attacks, and setting restrict permissions and access for different users.[22] In addition, there are many Linux security extension enhancements, such as ExecShield, and Position Independent Executable.[23]

Security is important for biomedical software tools because they are often used to manage and process patient data. To protect patient privacy, i2b2 provides secure remote access through web services and adds randomly generated noise to the count of a defined patient cohort in each institution.

Globus has maintained a strong security model for many years, using standards-based components and protocols that address message protection, authentication, delegation and authorization for distributed infrastructures. Globus authorization is based on well-established standards such as OAuth 2 and OpenID Connect and leverages federated login to allow user authentication using one of the many supported identity providers (e.g., institutional identities, eRA Commons, ORCID, and Google). The Globus high assurance tier provides additional security controls to meet the higher authentication and authorization standards required for access to restricted data, such as Protected Health Information. Data transfers can be encrypted using OpenSSL libraries, and communication channels with the Globus service are TLS 1.2 encrypted. Other OSS use cases have weaker security designs. Security enhancement is becoming the focus of future releases.

**Legal**

Among the ten OSS use cases, one popular licensing model is GNU General Public License (GPL), which allows the distribution and sale of modified and unmodified versions but requires that all those copies be released under the same license and be accompanied by the complete corresponding source code. For example, Linux was released under GPL v2; R and tranSMART used GPL v3.

Two use cases have mixed licenses, with GPL for some packages and Berkeley Software Distribution (BSD) for other packages. For example, Bioconductor packages belong to multiple license groups: Artistic license v2 (commercially friendly); GPL (distribution of any modified code must make the source available); and MIT, BSD, Creative Commons (permissive licenses that have minimal requirements about how the software can be redistributed).

Globus also uses mixed licensing models: (1) client-side software is licensed under the Globus Community License, which allows subscribers to access source code for the purposes of code review and contribution; (2) software operated by Globus as a service is not licensed.

Open source licensing models used by selected OSS use cases include Apache 2 (OHDSI, ITK) and Mozilla Public License version 2.0 under the terms of the Healthcare Disclaimer addendum (i2b2), and GPL3 (tranSMART). While REDCap requires a non-profit end-user license agreement to be made between an institution and Vanderbilt University and its code base is not open to an individual developer. Lastly, the 3D Slicer license, while generally being highly permissive, is not a standard Open Source Initiative certified license. Instead, it is a custom license that was defined via coordination with the legal department of Brigham and Women's Hospital, primarily aiming to mitigate liability risks due to the nature of the application (visualization and analysis in support of research applications on clinical images).

**Finance**

Most of the ten OSS use cases started with federal research funding. For example, Bioconductor began receiving NIH National Human Genome Research Institute support since 2003, and NCI/ITCR funding since 2014. 3D Slicer has had direct or indirect support from many research grants (primarily NIH) over the course of several decades, but no sustained funding from a single source or program. Cytoscape has support from the National Institute of General Medical Sciences and the National Resource for Network Biology. REDCap had early support from the National Center for Research Resources. Early Globus development was supported by the National Science Foundation and Department of Energy, and more recent work on high assurance mechanisms has been supported by the NIH.

Industry and membership support are common for mature OSS cases. For example, Premium Globus features (e.g., data sharing, usage reporting, guaranteed support levels) are offered to institutions under an annual subscription, which is a flat annual fee based on the institutions' level of research activity. Linux has been supported by individual memberships (over thousands of members) and corporate annual membership (over 1,000 corporate members). The R foundation is supported largely by members (membership fees from supporting persons, institutions, and benefactors) and "one-off donations."

Multiple sponsor programs involving both academic sponsors and industry sponsors are also feasible. For example, ITK has continual funding from NIH for maintenance to enable free use, and at the same time has commercial grade supports. OHDSI also has both private and public funding supports. The i2b2 tranSMART Foundation has four sponsorship programs: contributing sponsors, corporate sponsors, sustaining sponsors, and event sponsors. Through the tranSMART and the successor i2b2-tranSMART Foundation efforts, Keith Elliston and colleagues started Axiomedix in 2018 specifically to provide a commercial (for profit) support mechanism for government funded open source. Axiomedix offers a four-part business model that helps to support and sustain the open source platforms. First is a commercial grade software publishing and support model; second, a full service solutions offering; third, a software development and customization model (the Axiomedix Expert Network) that enables core open source developers to do contract and consulting for customers; and finally, a model for developing new products and platforms that leverages open source tools, an experience open source developer network, and the needs of subject matter experts to develop new open source or commercial tools.

**Marketing**

The ten OSS use cases have a variety of marketing channels, including logos (3D Slicer, Globus, i2b2 tranSMART), websites (3D Slicer, Bioconductor, Globus, i2b2 tranSMART), mailing lists (Cytoscape, Globus, i2b2 tranSMART), forums (3D Slicer, Cytoscape, i2b2 tranSMART), Twitter (3D Slicer, Bioconductor, Cytoscape, Globus, i2b2 tranSMART), LinkedIn (Globus, i2b2 tranSMART), Facebook (i2b2 tranSMART), YouTube (3D Slicer, Bioconductor, i2b2 tranSMART), Tumblr (Cytoscape), Vimeo (Cytoscape), and Pinterest accounts (Cytoscape).

A few OSS use cases mainly promote themselves through academic conferences, workshops and publications. For example, ITK is introduced at medical imaging conferences. R gains market share through an "evangelist" approach amongst statisticians, data analysts and others from the biomedical community.

Moreover, surveys are administered to collect users' feedback. For example, the 3D Slicer team conducts small-scale surveys on forum and collects feedback forms during training courses. Similarly, the Globus team conducts surveys during workshops and tutorials.

**Dependency hygiene**

Most of the ten OSS use cases have many dependencies on other packages. Bioconductor and OHDSI depend on many R packages. REDCap depends on MySQL. Cytoscape relies on some external services, including cxMate. Because dependencies may complicate installation and usage, i2b2 provides Docker containers for easy installation. Software models mainly provide dependency information through documentation, like installation guides, but few models describe the license and security status of each dependency.

**DISCUSSION**

We discussed ten representative OSS use cases that have demonstrated sustainable practices, particularly in the biomedical domain. Though not comprehensive, these use cases highlight the following essential attributes of successful OSS development: alignment with unmet scientific needs, dedicated development team, vibrant user community, feasible licensing model, and sustainable financial model.

**Alignment with unmet scientific needs**

At the inception of an OSS project, the project must identify and meet important scientific needs that no other suitable solution has met. This attribute is the "soul" of the software that gives it its identity. For example, Cytoscape fulfills the need for a visualization tool to represent complex interactions among molecules, Bioconductor reduces the barrier to entry involved in the effective use and share of computational biology and bioinformatics tools, and Globus addresses the need for frictionless data transfer and sharing.

Masys et al.[24] have discussed the importance of community needs. They pointed out that the adoption of an enterprise-level software tool should be highly motivated by users' immediate needs, not by mandatory rules or external financial rewards. In other words, filling existing scientific gaps should be a priority in an OSS project.

However, we should realize that scientific community needs are diverse and dynamic. Even at the initiation stage, developers should carefully consider the potential expansions beyond the first application and adopt an infrastructure with the highest reusability.

**Dedicated development team**

An OSS project should have a core development team who not only has developed an initial version of the software, but also continues to be committed to future versions. This attribute is the "brain" of the software that provides it its intellectual embodiment. For example, Globus includes services for identity management, data transfer, data sharing, and group management; interfaces such as APIs, web apps and a command line client; and software to manage data access on over ten distinct storage platforms and file systems. Only a dedicated and highly experienced development team can put all such components together in a concerted fashion.

According to Atiq et al.,[16] the motivations of developers usually include both intrinsic aspects (e.g., creativity, fun) and extrinsic aspects (e.g., financial rewards, development of job-related skills, and peer recognition). They further pointed out that transparent and fair extrinsic rewards and effective and open communications among developers are key characteristics to ensuring the long-term sustainability of OSS projects.

**Vibrant user community**

An OSS project should have a vibrant user community whose organizational structure and ongoing activities can facilitate communication (among developers, among users, and between developers and users) and "materialization" of the value of the software, while specifying the functionality requirements for future versions. This attribute represents the "heart" of the software that drives the development cycle. For example, 3D Slicer and ITK have large and stable user bases, mainly in the radiology and biomedical imaging communities. OHDSI has large user bases in clinical informatics and population health informatics communities. We highly recommend engaging scientists outside the original team and involving a broad array of stakeholders.

It's also important to realize that "users" of enterprise-level OSS are institutions, not individual researchers. Masys et al.[24] defined a successful adoption as at least 50% of the intended institutions adopting and implementing the tool. They suggested that, instead of a one-size-fits-all technical approach, developers should provide flexible local implementations and customizations (such as the optional use of terminology standards). This flexibility is essential to building a vibrant user community and reaching successful adoptions.

**Feasible licensing model**

A sustainable OSS project also needs a licensing model that fits the nature of the software, its distribution channel, and stakeholder interests. This attribute resembles the "skeletal system," providing a legal framework for the software to function properly.

Open source software licensing falls generally into four categories: fully permissive, weakly permissive, non-permissive, and non-compliant. Open source licenses are evaluated as conforming to the Open Source Definition by the Open Source Initiative, a 501c3 non-profit established to be a steward of open source licenses. Fully permissive licenses provide an unrestricted reuse of code for commercial and non-commercial purposes. Fully permissive licenses include Apache 2.0, MIT and BSD, among others. Weakly permissive licenses (MPL2 for example) allow commercial and non-commercial use but require release of any modified code on a file by file basis. Non-permissive licenses, such as GPL and AGPL, allow commercial and non-commercial reuse but also require the release of all modified code and any external code linked to this code. These non-permissive licenses are considered 'toxic open source licenses' as few if any commercial entities will reuse this code due to the extensive code release requirements of the license. Finally, many projects that are considered open source, release code under custom, non-OSI compliant licenses. Thus, even though they may make code available, these projects cannot be considered as open source compliant. There is a slow migration in the scientific software field to move toward fully permissive licenses, as few projects with non-permissive licenses can be sustained.

Elster also had an interesting discussion about how the license of scientific software may have an impact on obtaining industrial funding support.[25] Many informatics technology companies choose academic software with BSD licenses (full permissive licenses) over GPL licenses (non-permissive licenses), because GPL licenses add restrictions to code reuse in commercial software, raising concerns about future commercialization, while BSD licenses allow inclusion of code in commercial code. On the other hand, some companies prefer GPL licenses to BSD-like licenses, because they do not want their competitors to build commercial code on top of open source BSD software that they previously funded.

**Sustainable financial model**

Lastly, an OSS project requires a sustainable financial model (formal or informal) that can keep the software and its community moving forward. This attribute is a part of the "circulatory system," supplying "blood" to sustain the software ecosystem. The i2b2 tranSMART, Globus, and Linux are excellent examples that leverage multiple types of sources to sustain software development.

The public-private partnership is becoming a feasible way to support OSS project in the long-term, but the establishment of this partnership may not be easy. Industry partners usually have concerns about profitable commercialization time. An OSS project's public release of its knowledge and source code may allow market competitors to catch up quickly, while traditional commercialized software businesses usually conceal intellectual property as long as possible. On the other hand, an OSS project may quickly attract a large number of outside users and new developers, whose contributions can improve the robustness of a product, enabling platform-based customizations in multiple institutions. The robust implementations and large size of users will increase the commercial potential of the OSS project.

**Summary**

Software sustainability has been a persistent challenge recognized by initiatives such as the BD2K and the National Center for Data to Health (CD2H). BD2K[26] was a trans-NIH initiative launched in 2013 to support the research and development of innovative and transformative approaches and tools, which maximize and accelerate the integration of big data and data science into biomedical research. BD2K recognizes that software is a necessary part of any modern solution to biological problems. Representing the shared interest of the national CTSA consortium, the CD2H is particularly interested in sustainability strategies for data management infrastructure, which again inevitably involves the sustainability of software tools revolving around clinical data.

We, the ITCR SIP-WG, summarized a gamut of models for long-term software sustainability. Each approach has strengths and weaknesses. For example, community-based sustainance, including appropriate forking of branch-development efforts, is in many ways ideal in that it leverages the collective and continuous effort of entire communities. However, it might not be appropriate for niche but important areas of development; it might over-emphasize broad adoption rather than quality, novelty or significance; and it might not be able to leverage efforts that don't follow the same open-source licensing structure. Commercialization, such as adoption of software modules in clinical workstations, leverages a large pool of resources and software libraries, in addition to a direct path to a broad user base willing to pay for it. However, it is limited by proprietary restrictions and by its dependency on profit-making motives, which might not align well with biomedical significance or with "investment for the future" policies. Various infrastructure-based models can be effective ways to pool resources and avoid replication; however, they depend on a decision mechanism for the selection of the small percentage of software products that would be supported. Moreover, they might be less prone to supporting innovation, due to their not-so-dynamic nature. Various funding-based mechanisms combine the advantages of dynamic selection and evolution of software products, through the process of merit-based reviews, but they are limited by the harsh reality that existing funding is far less than the cost of long-term maintenance of meritorious software, a situation that is unlikely to change in the foreseeable future.

Looking forward, it will be important to engage with other groups interested in sustainable software models. One notable community is the Workshop on Sustainable Software for Science: Practice and Experiences (WSSSPE), a workshop series aimed at promoting sustainable research software by focusing on principles and best practices, careers, learning, and accreditation. The fourth WSSSPE[27] announced a group interested in writing white papers that focus on scientific environments and their implications, targeting at developers and project managers of scientific software.

**CONCLUSION**

Our review of existing sustainability models and ten OSS use cases strongly confirmed the importance of the three proposed future focuses in SIP-WG: 1) develop a workflow/decision tree to support informed decision making, consistent with ITCR expectations and the future licensing needs of open source tools; 2) provide a licensing consultancy service, in collaboration with the ITCR program; and 3) develop a collection of business model archetypes which can serve as starting templates, to formally document dissemination and sustainability plans for new software development initiatives.

In addition, we stressed five important actions to be considered in future ITCR activities:

**(1) Discuss the feasibility of sustainability models for ITCR projects**

One important agenda item of the SIP-WG future work should be discussing the feasibility of various sustainability models for many ITCR support projects, including nonprofit models (e.g., not-for-profit organization model, national funding model, infrastructure model, institutional support model, and donation model), and commercial models (e.g. distributed network model, content licensing/industrial support model, user subscription model, freemium model, split licensing model, razor and blades model, macro R&D infrastructure model, and mixed models).

**(2) Explore the potential licensing models**

The licensing of scientific software will have a direct impact on the private-public partnership. A mixed licensing model may be the best way to strike a balance between free usage (for broad use) and for-fee usage (for funding support). Given the potential complexities of different OSS approaches, key stakeholders should consider the license structure of their software models as early as possible. Important decisions and moves must align well with the roadmap of software development and maintenance, because changing the license of existing projects can be very challenging. Once an open source project integrates code from external contributors, it becomes logistically difficult to legally change the license on the code.

**(3) Provide reward mechanisms to enhance stakeholders' motivation on sustainability**

The WSSSPE community has pointed out the importance of enhancing stakeholders' motivation by credit and rewards.[28] Currently, the main credit given for developing an academic OSS are through publications. We should encourage key contributors to list the creation of software resources on their biosketches and further value OSS in the grant funding review process. We should also provide rewarding mechanisms in order to fairly allocate credit to external developers that have contributed to successful expansions and adoptions. Finally, universities and research institutions should create career paths with a bright future for scientific software engineers to encourage them to continuously work on academic OSS development.

(4) **Establish a central library to make OSS visible and reusable**

In addition, we should consider establishing a central library to make ITCR-funded OSS more visible and reusable for a large number of cancer researchers. The open-access library should index OSS tools with brief descriptions of functions and simple examples. This library will point to the latest version of each OSS tool. It will especially serve as a repository for retiring OSS tool, which may have short-term difficulties in getting funding support. Ideally, this library should be searchable, like a Google for academic OSS. When researchers have certain needs, they can first search within this library to find out whether there is an existing tool available to meet their needs or whether there is an existing tool that they may expand to meet their needs.

**(5) Provide industry standard support on code quality control, ecosystem collaboration, security, and dependency hygiene**

Finally, we should allocate consulting resources to academic OSS projects (especially at the early stage development), which can guide these projects to follow up-to-date industry standards on code quality control, ecosystem collaboration, security, and dependency hygiene.

**Open Source Software Sustainability Models: Initial White Paper from the Informatics Technology for Cancer Research Sustainability and Industry Partnership Work Group (Supplementary file)**

Ye Y (U Pitt), Boyce RD (U Pitt), Davis MK (U Pitt), Elliston K (tranSMART Foundation & Axiomedix, Inc.), Davatzikos C (U Penn), Fedorov A (Harvard), Fillion-Robin JC (Kitware Inc.), Foster I (U Chicago), Gilbertson J (U Pitt), Heiskanen M (NCI ITCR), Klemm J (NCI ITCR), Lasso A (Queen's University), Miller JV (GE Research), Morgan M (RPCI), Pieper S (Isomics Inc.), Raumann B (U Chicago), Sarachan B (GE Research), Savova G (Harvard), Silverstein JC (U Pitt), Taylor D (U Pitt), Zelnis J (U Pitt), Zhang GQ (UT Health) and Becich MJ (U Pitt).

**Table 2. Summary Comparison of Governance of Academic OSS**

| Software tool | Characteristics |
|---|---|
| **3D Slicer** | • Decision making is typically happening on the GitHub pull requests, user/developer hangouts and in-person meetings; decisions are publicly documented via GitHub, forum discussions, and Wiki pages; Roadmap for the development is publicly available.<br>• Contribution guidelines are publicly available; core developers and users meet in person twice a year; sub-communities have regular meetings.<br>• Feasibility studies and experimental designs are documented on "Labs" pages.<br>• Forum and video conference meetings are publicly available; source code is in GitHub; bug tracker is publicly available; training courses are organized at major conferences and educational institutions. |
| **Bioconductor** | • Technical advisory board meets monthly to develop strategies to ensure long-term technical suitability of core infrastructure and to identify funding strategies for long-term viability.<br>• Scientific advisory board includes external experts, providing annual guidance. |
| **Cytoscape** | • Cytoscape is architected as core software augmented by Cytoscape Apps.<br>• Core team contributes to the core software through GitHub commit.<br>• Cytoscape 3.0 has a clearly defined, simplified API.<br>• Each class in the public API maintains an explicit backwards compatibility contract so that both core developers and app writers understand how this class might change. |
| **Globus** | • Technical and business decisions are in the hands of the Globus team.<br>• Institutional subscribers via subscriber meetings and consultations, and community users via mail lists, webinars and on-site tutorials, provide input on feature prioritization, product roadmap, subscription pricing, and sustainability models. |
| **i2b2 tranSMART** | • The i2b2 tranSMART Foundation was established; the bylaws are available publicly.<br>• The foundation includes high-level leaders, membership program, and working groups, communicating through mail list and recorded online webinars. |
| **ITK** | • ITK consortium manages licensing, owns the copyright, and acts as a governing body.<br>• Code changes are strictly controlled by limiting the commit of repository changes.<br>• A code review process has been established for submissions.<br>• As a forum for proposed changes, an online ITK journal documents changes and acts.<br>• Development is maintained fully via GitHub.<br>• Contribution guidelines are documented. |
| **Linux** | • Linux is the premier example of open source sustainability and success.<br>• Linux Foundation was founded in 2000 as a neutral home, providing fellowships, IT operations, training, and events.<br>• Board of directors is comprised of 22 senior leaders from across the IT industry.<br>• Board members represent Linux Foundation members and the Linux developer community. |
| **OHDSI** | • OHDSI has an international network, centrally coordinated at Columbia University.<br>• OHDSI hosts a collection of open source tools, among which OMOP related standard vocabulary are foundational tools. |
| **R** | • Since mid-1997, there has been a core team responsible for overseeing its development.<br>• R Foundation, a registered association under Austrian law, is currently hosted by the Vienna University of Economics and Business; R Foundation supports the development, teaching and training, and the organization of meetings and conferences.<br>• Ordinary members of the Foundation are elected by a majority vote of the general assembly.<br>• New ordinary members are selected based on their non-monetary contributions. |
| **REDCap** | • Consortium consists of non-profit organizations interested in expanding functionality.<br>• Each partner institutional is hosting as a "platform" to reach out and bring value to its intended end users; site was given access to the codebase; code is available at no charge to institutional partners and is restricted in use, permitted only for non-commercial research purposes.<br>• Vanderbilt is the only entity that can distribute it; any and all derived works – such as innovations or programmatic features added on by the user – are owned by Vanderbilt. |

**Table 3. Summary Comparison of Documentation of Academic OSS**

| Software tool | Characteristics |
|---|---|
| **3D Slicer** | • Documentations include Wiki, ReadTheDocs, crowd-sourced documentation, various recipes, and YouTube videos.<br>• Tutorials page with slides and sample datasets materials are publicly available.<br>• Commit style guidelines are documented publicly, with human-focused readable notes.<br>• It does not have support for internationalization of the documentation, but there are user communities that developed documentation in their languages independently, e.g., in Chinese. |
| **Bioconductor** | • Three-level documentation is available: workflows, package vignettes, and function manual pages.<br>• Users often become developers, making their own algorithms and approaches available to others. |
| **Cytoscape** | • Extensive user and developer documentation: developer resources include an issue tracker, Nexus repository, nightly builds, and code metrics.<br>• Cystoscape "App Ladder" teaches essential skills for developing Cytoscape Apps.<br>• Publicly available resources include basic and advanced tutorials, a YouTube channel, a blog of published figures, as well as several presentations. |
| **Globus** | • Fully developed documentation is available, including installation guides, API usage guides, FAQs, tutorials, and how to guides.<br>• Each release is clearly versioned with release notes and change history. |
| **i2b2 tranSMART** | • i2b2 updates documentation through a Wiki page; developers commit messages in GitHub; i2b2 provides human-focused notes of changes of each release and bug fixes.<br>• tranSMART provides documentation on version 16.3. |
| **ITK** | • ITK provide a user guidebook.<br>• Coding examples are automatically built as part of nightly tests. |
| **Linux** | • Code is available through GitHub. |
| **OHDSI** | • Available documentation introduces how to get started with OHDSI, common Data Model, and ETL creation best practices, as well as tool specific information. |
| **R** | • R provides various fully developed documentation, adequately covering two types of development: writing R extensions and developing R itself (by providing internal structure and coding standards). |
| **REDCap** | • Detailed documentation is available for setting up and software usage, but not on contributing source code. |

Table 4. Summary Comparison of Code Quality of Academic OSS

| Software tool | Characteristics |
|---|---|
| **3D Slicer** | • Code styling guidelines are documented publicly; new functionality is contributed by modifying the core application or by submitting an extension (code styling guidelines are not enforced for extensions).<br>• Developers conduct about 700 tests for the core application and testing results are available publicly; extension contributors manage the test for extensions (not verified by core developers).<br>• Contribution process is documented publicly. |
| **Bioconductor** | • Packages are built nightly on Windows, macOS, and Linux platforms.<br>• Packages are built successfully before propagating to public user-facing repositories.<br>• Tests are conducted to ensure overall package integrity and integration with current versions of package dependencies. |
| **Cytoscape** | • Core developers use Jenkins and open source automation tool to build and test software projects continuously.<br>• Contributors on external extensions are strongly encouraged to thoroughly test apps before releasing them; the Cytoscape team does not independently review these apps. |
| **Globus** | • Professionally maintained and managed code base employs best practices, such as revision control in GitHub, extensive documentation, code comments, linting, distribution of knowledge across team members, and incremental/iterative software development.<br>• Globus employs a continuous integration environment, automated tests, and documented, standardized human QA testing to ensure code quality.<br>• Code is reviewed by at least one engineer other than the code author before being released to production.<br>• Globus developers follow secure development practices, including OWASP recommendations to prevent web application security risks. |
| **i2b2 tranSMART** | • Both i2b2 and tranSMART have extensive automated and manual testing as a part of their well define release process. |
| **ITK** | • Automated nightly builds and tests are as far back as 1999.<br>• Strictly enforced coding conventions with consistent naming rules. |
| **Linux** | • The code quality has not been discussed in case profiling. |
| **OHDSI** | • The code quality varies, because a large collection of tools is developed around the OMOP CDM. |
| **R** | • Guidelines are maintained for its Software Development Life Cycle; coding standards are established.<br>• Apache Subversion is used to maintain current and historical versions of files; software development and testing methodologies are employed by R Core in order to maximize the accuracy, reliability, and consistency; some aspects are handled collaboratively, others are handled by members of the team with specific interests and expertise in focused areas.<br>• R can run on a wide variety of UNIX platforms and similar systems (including FreeBSD and Linux), Windows and MacOS. |
| **REDCap** | • Source code does not open to the community. |

Table 5.  Summary Comparison of Support of Academic OSS

| Software tool | Characteristics |
|---|---|
| **3D Slicer** | • Support is provided by the community of developers and users; only a small fraction has resources for 3D Slicer development; even smaller fraction has dedicated funding.<br>• For the over 13,000 forum posts in 2018, the average response time was less than 2 days (or less than 8 hours during weekdays).<br>• Active members on several forums and mailing lists provide support via various social media platforms (e.g., Twitter, Research Gate, YouTube). |
| **Bioconductor** | • Individual "landing pages" (e.g., DESeq2) provides an overview of the package, installation instructions, and usage statistics.<br>• User-oriented stack overflow-style support site is publicly available and active (100's of visitors per hour; fast response).<br>• All packages include maintainers' email addresses and links to GitHub issues.<br>• Developer supports are available by the email lists.<br>• Hybrid community slack is available for experienced user / developer collaboration. |
| **Cytoscape** | • Cytoscape Help Desk |
| **Globus** | • Several support options are available, such as online self-help tools, listserv groups, and a ticket submission system with a responsive, dedicated support team.<br>• Guarantees subscribers are guaranteed a response time of one business day to support tickets. |
| **i2b2 tranSMART** | i2b2:<br>• bug tracker<br>• Google forum for installation help<br>tranSMART:<br>• tranSMART Wiki |
| **ITK** | • ITK has mailing lists; over its long history, ITK has had dedicated volunteers who would give detailed help to users even including example code.<br>• ITK has discourse forum for discussions and mutual help among users.<br>• NIH has continued to provide maintenance contracts for bug fixes, incremental improvements, and a moderate level of user support.<br>• Maintenance has typically been performed by Kitware, providing continuity and expertise. Kitware also offers commercial ITK support. |
| **Linux** | • Linux support is provided through the LF JIRA. |
| **OHDSI** | • Community-based discourse forum<br>• GitHub issue tickets |
| **R** | • Online self-help tools;<br>• FAQ listings<br>• Subscription-based email lists, including general R-help email list, the R-developer list, and R- package-developer list |
| **REDCap** | • REDCap is supported by Vanderbilt University. |

**Table 6. Summary Comparison of Ecosystem Collaboration of Academic OSS**

| Software tool | Characteristics |
|---|---|
| **3D Slicer** | • 3D Slicer leverages the ecosystem of related open source projects.<br>• More extensive list of dependencies is available publicly; developers interact on a regular basis with developers for most of the upstream dependencies and actively contribute bug fixes and feature improvements to the appropriate repositories/<br>• 3D Slicer has automated testing on all major platforms available to minimize the possibility of regressions and conflicts. |
| **Bioconductor** | • 1649 packages; over 1000 maintainers; over 29,000 mentions in literature; 500,000 unique IP addresses in 2018 |
| **Cytoscape** | • an active and diverse community of developers and users<br>• Communication happens mostly via mailing lists and social media (Twitter, publications Tumblr). |
| **Globus** | • Large user base of over 20,000 users per year and 140 institutional subscribers.<br>• Annual conference for users and subscribers.<br>• Active community listserv groups for users, admins and developers.<br>• Numerous webinars and on-site, two-day workshops occur every year at research institutions across the US.<br>• Many grant-funded collaborations to apply, extend, develop, or innovate Globus services in the context of science drivers.<br>• Professional services and customer engagement teams to enable collaborations and to facilitate application of Globus services to research data management needs. |
| **i2b2 tranSMART** | • Guidelines of submitting new contributions are publicly available. |
| **ITK** | • Over 200 developers contributing to the source code of ITK |
| **Linux** | • A strong and vibrant community: 13,594 developers from at least 1,340 companies have contributed to the Linux kernel since 2005; >1,600 developers contributed to each kernel version<br>• A new major kernel release occurs every 9–10 weeks; kernel community merged changes at an average rate of 7.8 patches per hour over the past 15 months.<br>• A thriving commercial ecosystem: 95%+ of the top 1M web domains, 80%+ of smartphones run Android, 98%+ of the top 500 fastest supercomputers in the world, most of the global markets, including NYSE, NASDAQ, London Exchange, Tokyo Stock Exchange.<br>• Most consumer electronic devices run on Linux; 75%+ of cloud-enabled enterprises report is using Linux as their primary cloud platform; E-commerce giants Amazon, EBay, PayPal, Walmart, and more run on Linux. |
| **OHDSI** | • Consortium is organized by projects and workgroups; OHDSI has annual symposium and satellite events, and community forum. |
| **R** | • two conference series: useR! (a user forum) and DSC (a developer platform)<br>• R core team meets collectively and/or in smaller groups |
| **REDCap** | • Consortium has two tiers of users with distinct characteristics: institutional users and investigator users.<br>• REDCap can be installed in a variety of environments for compliance with such standards as HIPAA, 21 CFR Part 11, and FISMA (low, moderate, high). |

Table 7. Summary Comparison of Security of Academic OSS

| Software tool | Characteristics |
|---|---|
| **3D Slicer** | • Online resources are available via HTTPS; downloads are available via HTTPS.<br>• No policy is used to enforce two-factor authentication for contributors.<br>• Binary packages are built on highly secured workstations, managed by Kitware staff, and protected by strict multi-factor authentication.<br>• 3D Slicer does not have specific consideration for addressing security aspects. |
| **Bioconductor** | • Security has not been discussed in case profiling. |
| **Cytoscape** | • A platform-independent open-source Java application is released under the terms of the LGPL. |
| **Globus** | • Globus has maintained a strong security model for many years, using standards-based components and protocols that address message protection, authentication, delegation and authorization for distributed infrastructures.<br>• Identity and access management is based on OAuth 2, providing an advanced, user consent based delegated authorization model that allow applications and services to act on behalf of users and other services.<br>• Encrypted data transfers use the SSL cipher configured on the source and destination storage systems. Communication channels with the Globus service are TLS 1.2 encrypted.<br>• Globus developers follow secure development practices, including OWASP recommendations to prevent web application security risks. Globus meets higher security standards for access to restricted data, including HIPAA regulated data.<br>• Security reviews by various external bodies, including code reviews by the NSF Trusted CI center and HIPAA risk and gap analyses by third party auditors. |
| **i2b2 tranSMART** | i2b2:<br>• Securable remote access is available through web services; random numbers are added to aggregate counts to protect privacy.<br>• Security enhancement is becoming the focus of the next core release.<br>tranSMART:<br>• Protected study operation requires administrator approval. |
| **ITK** | • Security is not a primary consideration for ITK, which is typically embedded within broader systems.<br>• ITK is not particularly vulnerable to cyber-threats. Being implemented in a low-level language (C++), some types of memory hacks may be possible. |
| **Linux** | • Linux has strong security features.<br>• The Linux kernel allows administrators to improve security at the lowest level by modifying attributes of the kernel's operation, building additional security measures into the kernel to avoid common buffer overflow attacks, and setting restrict permissions and access for different users.<br>• In addition, there are many Linux security extension enhancements, such as ExecShield, and Position Independent Executable. |
| **OHDSI** | • Regular discussions are related to toolset security.<br>• Apache SHIRO is used for securing the WebAPI component. |
| **R** | • An inherent security risk in malicious code is making their way unnoticed into the central repositories; there are a few published papers on the potential security risk of R and general guidelines on how to handle security issues.<br>• No definitive information on security for privately reporting security issues<br>• No provenance chain for releases compiled binaries<br>• No reported vulnerabilities produced for security issues published within a reasonable timeframe |
| **REDCap** | • It has a secure web connection with authentication and data logging. |

**Table 8. Summary Comparison of Legal of Academic OSS**

| Software tool | Characteristics |
|---|---|
| **3D Slicer** | • Non-restrictive (commercial use permitted) license<br>• Not an OSI-approved license<br>• Specific license is defined via coordination with the BWH legal department, primarily aiming to mitigate liability risks. |
| **Bioconductor** | • Almost all Bioconductor packages are licensed in a way that allows use by any entity without permission.<br>• A small number of packages specify "academic only" use.<br>• Bioconductor packages belong to multiple license groups: Artistic license v2 (commercially friendly); GPL (distribution of any modified code must make the source available); and MIT, BSD, Creative Commons (permissive licenses that have minimal requirements about how the software can be redistributed). |
| **Cytoscape** | • Consortium holds the intellectual property rights of Cytoscape (LGPL), Cytoscape.js (LGPL), NeXO (BSD or similar, in progress), and NDEx (BSD or similar, in progress). |
| **Globus** | • Globus uses mixed licensing models.<br>• Select software components operated as a service and client-side software, such as the Globus Command Line Interface, the Globus Software Development Kit, and some versions of Globus Connect (the software that enables storage systems to be accessed via the Globus service) are made available under the Apache 2.0 license.<br>• Some versions of Globus Connect are licensed under the Globus Community License, under which subscribers can access source code for the purposes of code review and contribution.<br>• The remaining software components operated as a service by Globus are not licensed. Globus holds no patents on Globus technology. |
| **i2b2 tranSMART** | • The i2b2 software is licensed through the Mozilla Public License (MPL) version 2.0 under the terms of the Healthcare Disclaimer addendum.<br>• The tranSMART software is available under the terms of the version 3 of the GNU General Public License. |
| **ITK** | • ITK is licensed under Apache that requires contributors to abide by the open source requirements.<br>• ITK does not have no embedded third-party packages that have licensing restrictions.<br>• Patented materials were no longer permitted for inclusion. |
| **Linux** | • Linux is released under GNU General Public License (GPL) v2, which allows distribution and sale of possibly modified and unmodified versions but requires that all those copies be released under the same license and be accompanied by the complete corresponding source code. |
| **OHDSI** | • OHDSI packages have open Source licenses, such as Apache License 2. |
| **R** | • R is under the terms of the Free Software Foundation's GNU General Public License. |
| **REDCap** | • Non-profit End-User License Agreement is made between an institution and Vanderbilt University ("Vanderbilt"), a not-for-profit corporation duly organized and existing under the laws of Tennessee. |

Table 9. Summary Comparison of Finance of Academic OSS

| Software tool | Characteristics |
|---|---|
| **3D Slicer** | • No dedicated organization<br>• 3D slicer is supported by academic groups and commercial entities, either choosing voluntarily to contribute resources to 3D Slicer development or applying/allocating research or product development funding.<br>• Numerous research grants (primarily NIH)<br>• Enforcing disclosure of financial support would not be desirable, as commercial entities often do not want to publicly acknowledge what open-source software they use in their products, or development of what features they actively support. |
| **Bioconductor** | • Initial support was from institutional funding.<br>• NIH/NHGRI support began in 2003<br>• NCI/ITCR funding began in 2014;<br>• Additional grants included the NSF, European Union, NCI ITCR, NHGRI, and the Silicon Valley Foundation.<br>• Details are available in annual reports<br>• Funding sources are almost never acknowledged in contributed packages. |
| **Cytoscape** | • Funding for continued development and maintenance of Cytoscape is provided by the U.S. National Institute of General Medical Sciences.<br>• User support, education and new initiatives are supported by the National Resource for Network Biology. |
| **Globus** | • Globus is provided to the community via a freemium model.<br>• Researchers may use Globus services for free.<br>• Institutions pay a flat, annual subscription fee, based on the institution's level of research activity, for unlimited use of premium features. |
| **i2b2 tranSMART** | • The foundation has four sponsorship programs: contributing sponsors, corporate sponsors, sustaining sponsors, event sponsors<br>• Through the tranSMART and the successor i2b2-tranSMART Foundation efforts, Keith Elliston and colleagues started Axiomedix in 2018 specifically to provide a commercial (for profit) support mechanism for government funded open source. |
| **ITK** | • ITK is free to users.<br>• NIH has provided continual funding for maintenance.<br>• Commercial-grade support can be purchased. |
| **Linux** | • Foundation is a membership organization based on different levels of membership: individual memberships (over thousands), corporate annual membership (over 1,000 corporate members) |
| **OHDSI** | • OHDSI is not specifically funded by any single organization.<br>• Funding sources have included private and public sources. |
| **R** | • R is funded largely by supporting members and "one-off donations."<br>• Annual membership fees are gathered from supporting persons, institutions, and benefactors. |
| **REDCap** | • Early support was provided by NCRR. |

Table 10. Summary Comparison of Marketing of Academic OSS

| Software tool | Characteristics |
|---|---|
| 3D Slicer | • Logo, web site, publications, YouTube channel, and Twitter<br>• There is no "marketing strategy" or dedicated personnel to support project marketing.<br>• Official announcements of releases and major events are regularly made via forum.<br>• Small-scale surveys are conducted on forum to get general feedback.<br>• Information is also collected via feedback forms and during training courses and thematic discussions during project weeks. |
| Bioconductor | • A consistent web presence and annual conferences<br>• Social media (e.g., twitter) and video sharing (YouTube) channels |
| Cytoscape | • Marketed through Twitter, Tumblr, Vimeo, and Pinterest<br>• Mailing lists and forums for specialized audiences. |
| Globus | • Memorable name, unique logo, public website optimized for search<br>• Participation at industry events<br>• Social media (Twitter and LinkedIn)<br>• News media via occasional press releases and article postings<br>• Blog with frequent announcements and articles;<br>• Email announcements, public workshops and user conferences, annual user conference<br>• Enthusiastic user evangelists<br>• Partnered with several commercial and non-commercial organizations |
| i2b2 tranSMART | • Logo, trademark<br>• Official announcements available through official website<br>• Workgroup mail lists<br>• Administrator and developer mail lists<br>• Social media (Twitter, LinkedIn, Facebook, Google forum; YouTube) |
| ITK | • Promotion at medical imaging conferences |
| Linux | • As of March 2016, top corporate contributors to Linux kernel include Intel, Red Hat, IBM, Motorola, Linaro, Google, Mellanox, SUSE, AMD, Renesas Electronics, Samsung, Rockchip, Oracle, ARM, Canonical, Broadcom.<br>• The Linux system is available as a product from Red Hat, which provides storage, operating system platforms, middleware, applications, management products, and support, training, and consulting services. |
| OHDSI | • Not a business entity<br>• Not actively engage in marketing<br>• Specific workgroup is focusing on dissemination.<br>• Researchers provide acknowledgment to OHDSI in publications and presentations. |
| R | • No direct information on marketing strategies<br>• R gained market share by an "evangelist" approach amongst statisticians, data analysts and others from the biomedical community.<br>• Some major commercial software systems are supporting connections to or integration with R, such as MATLAB, SAS, etc. |
| REDCap | • Marketing is not discussed in case profiling. |

**Table 11. Summary Comparison of Dependency Hygiene of Academic OSS**

| Software tool | Characteristics |
|---|---|
| **3D Slicer** | • Compatibility of the dependencies is considered and checked by means of automated testing.<br>• No established practice of checking potential security issues of dependencies.<br>• The entire 3D Slicer code base is built from source every night on Windows, macOS, and Linux; build results are posted to a public dashboard that is cross-linked to the corresponding source code repositories. |
| **Bioconductor** | • Dependencies on other R packages are from either Bioconductor or CRAN.<br>• Packages rarely depend on third-party software, or software written in other languages.<br>• Deeper dependency graph arises in part from emphasis on re-use of robust and interoperable core- developed infrastructure.<br>• Some packages try to capture A-to-Z 'workflows' and thus tend to have a large number of dependencies, these tend to be fragile. |
| **Cytoscape** | • Functions rely on code deployed as services available on web servers.<br>• Generally, such services are callable by Cytoscape or directly by non-Cytoscape clients (e.g., Python) in the larger bioinformatics community.<br>• Some services are provided by other organizations and located in or rely on other GitHub repositories.<br>• Known external repositories containing services that are called by Cytoscape include CXMate, Diffusio. |
| **Globus** | • Transition to a SaaS model has greatly reduced software dependencies.<br>• Only Python libraries for Globus Connect Server, and OpenSSL and SSH for Globus Connect Personal, are included in shipped software, which are generally updated with each software release.<br>• GitHub alerts are received for any security issues for dependent packages.<br>• Dependencies that are not included in Globus software packages are documented in installation instructions. |
| **i2b2 tranSMART** | • Installation guide lists dependencies.<br>• Docker containers are available for easy installation.<br>• Each release is analyzed for dependencies, and to ensure that there is a definitive list of components within the system. |
| **ITK** | • A name-mangling scheme is preventing third- party packages from having name conflicts.<br>• It is easy for users to build.<br>• It does not have a built-in package manager, such as pip for Python. |
| **Linux** | • The Linux kernel has a few dependencies, which are publicly listed. |
| **OHDSI** | • OHDSI depends on individual tools. Some uses R.<br>• WebAPI uses Maven for dependency management. |
| **R** | • Package is self-contained in general.<br>• Dependencies rely on the contributions from its community of developer libraries within the central CRAN.<br>• Distribution is under the terms of the Free Software Foundation's GNU<br>• Package dependencies are clearly document within the package summary documentation within the CRAN site. |
| **REDCap** | • Web server with PHP, Apache (any OS) or Microsoft IIS (Windows)<br>• MySQL database server<br>• MySQL client<br>• SMTP email server |

# Open Source Software Sustainability Models (OSSSM): Use Cases

## Use Case: 3D Slicer

https://slicer.org

**Author of Use Case**: Andrey Fedorov (BWH); Contributors: Steve Pieper, Jean-Christophe Fillion-Robin, Andras Lasso.

**Short History of Tool**: 3D Slicer is a free open-source extensible platform for medical image computing and visualization. Slicer started as a software tool to support planning and visualization for image-guided neurosurgery developed by David Gering as part of his Master's work at MIT (Gering et al. 1999; Fedorov et al. 2012). Since 1999, Slicer has been under continuous development at the SPL under the leadership of Ron Kikinis. Today it is developed mostly by professional engineers in close collaboration with algorithm developers and application domain scientists, with the participation of Isomics Inc., Kitware Inc., GE Global Research and Queen's University, and with significant contributions from the growing Slicer community. Initially envisioned as a neurosurgical guidance, visualization and analysis system, over the last decade, Slicer has evolved into an integrated platform that has been applied in a variety of clinical and preclinical research applications, as well as for the analysis of non-medical images (Fedorov et al. 2012) as well as a more recent and dynamically updated list of use cases and users here: https://www.slicer.org/wiki/Main_Page/SlicerCommunity).

**Who or what is 3D Slicer competition?**

3D Slicer is a standalone desktop application that targets primarily medical imaging researchers, and generally anyone with the needs related to volumetric image visualization, analysis and development of customized application interfaces. For the analysis with the tools that provide similar functionality the reader is referred to (Fedorov et al. 2012).

**Governance:** The source code is hosted under open GitHub organization: https://github.com/Slicer. Contribution guidelines are formalized in publicly available document: https://github.com/Slicer/Slicer/blob/master/CONTRIBUTING.md, which includes pointers to the developer documentation, formalized decision making process and governance structure. Slicer forum is open for anyone interested in participating, with archives publicly available and searchable: https://discourse.slicer.org. Open video conferencing meetings are taking place weekly, with anyone welcomed to join to ask questions or discuss related topics: https://discourse.slicer.org/c/community/hangout. Bug tracker is available publicly: https://issues.slicer.org/.

Slicer core developers and users meet in person twice a year (NA-MIC Project weeks/hackathons) to work on selected software development topics (application of the platform for specific research and development projects, feasibility studies, integration tests, etc.), provide training, have discussions, and make strategic decisions. These events have been continuously running since 2005, with typical attendance of 40-80 people, at various locations around the world: https://na- mic.github.io/ProjectWeek/. There are several sub-communities, each gathered around a specific Slicer extension (SlicerIGT, SlicerRT, SlicerSALT, SlicerHeart, etc.), with their own regular meetings, repositories, issue trackers, etc.

Decision making is typically happening on the GitHub pull requests, user/developer hangouts and in- person meetings. Decisions are publicly documented via GitHub, forum discussions and Wiki pages. Feasibility studies, experimental designs are documented on "Labs" pages (https://www.slicer.org/wiki/Documentation/Labs).

As illustrated by the history of pull requests associated with the GitHub project hosting the source code (https://github.com/Slicer/Slicer/pulls?q=is%3Apr+is%3Aclosed, 1066 total as of writing this), there are numerous developers contributing to the development of 3D Slicer. About 10 developers have permission to make modifications to the Slicer core directly, thus it is not relying on a single person. Most extensions are hosted on GitHub and maintained by small developer groups.

Roadmap for the development is available publicly and documented: https://issues.slicer.org/roadmap_page.php

Slicer forum, weekly video conferences, GitHub community, in-person bi-annual meetings, training courses organized at major conferences and educational institutions aim to create a diverse and welcoming community.

**Documentation:** Various sources of documentation are available, including Wiki, ReadTheDocs, crowd-sourced documentation, various recipes and YouTube videos. A challenge in managing documentation consistently is due to the extended history of the project (1999 was before GitHub and YouTube!)

Documentation has been provided originally using Wiki (both for users and developers): https://www.slicer.org/wiki/Documentation/Nightly. A relatively new effort is to migrate this documentation to the more modern and more usable ReadTheDocs platform: https://slicer.readthedocs.io/en/latest/index.html. API documentation is available at http://apidocs.slicer.org/.

Tutorials page with slides and sample datasets materials are available https://www.slicer.org/wiki/Documentation/Nightly/Training. Non-curated videos on YouTube educating on the use of Slicer: https://www.youtube.com/results?search_query=3d+slicer&sm=3

Commit style guidelines are documented publicly: https://www.slicer.org/wiki/Documentation/Nightly/Developers/Style_Guide#Commits, and are consistent (at least, for the most part) with the Chris' great guidance (see commit history is available https://github.com/Slicer/Slicer/commits/master).

Human-focused readable release notes are posted in the dedicated section of the Slicer forum: https://discourse.slicer.org/c/announcements/release-notes

Slicer funded efforts do not have support for internationalization of the documentation, but there are user communities that developed documentation in their languages independently, e.g., in Chinese.

Scientific publications and other work based on 3D Slicer for over a decade is collected on a Wiki as a reference for people to use when carrying out their use of 3D Slicer: https://www.slicer.org/wiki/Main_Page/SlicerCommunity/2019

**Code Quality:** The project does have code styling guidelines documented publicly: https://www.slicer.org/wiki/Documentation/Nightly/Developers/Style_Guide

New functionality can be contributed by modifying the core application, or by contributing an extension (similar to applications available in a mobile platforms app stores), the latter approach being recommended and most practical. Code style guidelines are not enforced or verified for the extensions, delegating that responsibility to the extension contributors.

The project does have testing integrated, as of writing, there is a total of around 700 tests for the core application. Tests for the extensions are managed by the extension contributors. Testing results are available on a public dashboard: http://slicer.cdash.org/index.php?project=SlicerPreview.

Contribution process is documented publicly in https://github.com/Slicer/Slicer/blob/master/CONTRIBUTING.md. Code review process is enabled by GitHub pull requests. Pull requests are tested using continuous integration.

Entry on BlackDuck/OpenHub with additional metrics: https://www.openhub.net/p/slicer

**Support:** Support is provided by the community of developers and users. Only a small fraction of the community has resources for 3D Slicer development. Even smaller fraction (if anyone at all) has dedicated funding for providing support. The main source of support is https://discourse.slicer.org

In spite of the voluntary nature of the 3D Slicer support forum activity, for the over 13,000 forum posts in 2018 the average response time was less than 2 days (or less than 8 hours during weekdays).

Community members are active on several forums and mailing lists where potential users and developers participate (for example, ITK, VTK, CTK forums) and monitor/provide support via various social media platforms (Twitter, ResearchGate, YouTube, Quora, etc.).

**Ecosystem Collaboration:** 3D Slicer is leveraging the ecosystem of related open source projects, which most notably include: VTK, ITK, Qt, CMake, Python, DCMTK. A more extensive list of dependencies is available at https://github.com/Slicer/Slicer/tree/master/SuperBuild. Automated testing on all major platforms is utilized to minimize the possibility of regressions and conflicts.

Slicer developers interact on a regular basis with developers for most of the upstream dependencies and actively contribute bug fixes and feature improvements to the appropriate repositories in preference to making local forks.

**Security:** Release binary packages are signed. Most online resources are available via HTTPS. Downloads of the binaries are available via HTTPS. There is no policy to enforce two-factor authentication for contributors. Binary packages are built on highly secured workstations, managed by Kitware staff, protected by strict multi-factor authentication. There is no specific consideration for addressing security aspects, considering the nature of the project.

**Legal:** The project is made available under a non-restrictive (commercial use permitted) license, which, however, is not an OSI-approved license. The specific license used was defined via coordination with the BWH legal department, primarily aiming to mitigate liability risks. License is available as part of the source code repository: https://github.com/Slicer/Slicer/blob/master/License.txt

**Finance:** There is no dedicated organization, foundation, or like that is responsible for supporting 3D Slicer development.

Maintenance and development are supported by academic groups and commercial entities either choosing voluntarily to contribute resources to 3D Slicer development, or applying/allocating research or product development funding.

Throughout the lifetime of the project, numerous research grants (primarily NIH) directly or indirectly supported development of 3D Slicer (nearly 50, as of writing). Documented list is available here: https://www.slicer.org/wiki/Documentation/4.x/Acknowledgments.

Note that the accounting above does not differentiate between supports for the development of the core application vs extensions. There are no mechanisms in place that would facilitate collection of detailed statistics or acknowledging support for the contributed extensions.

Enforcing disclosure of financial support would not be desirable, as commercial entities often do not want to publicly acknowledge what open-source software they use in their products, or development of what particular features they actively support.

**Marketing:** Project has a logo, web site, publications, forum, bug tracker, GitHub community, which aim to support the community and improve marketing of the project. However, there is no "marketing strategy" or dedicated personnel to support project marketing.

Official announcements of releases and major events are regularly made via Slicer forum ("Announcements" category). Interesting projects, developments, related efforts are posted on Twitter (https://twitter.com/3DSlicerApp). 3D Slicer YouTube channel is used to organize (mostly, community-contributed) educational videos about Slicer.

Small-scale surveys occasionally conducted on the Slicer forum to get general feedback from the community. Information is also collected from the community via feedback forms and during training courses and thematic discussions during project weeks.

**Dependency Hygiene:** There is no established practice of checking potential security issues. Compatibilities are checked by means of automated testing. The entire 3D Slicer code base is built from source every night on Windows, macOS, and Linux. Build results are posted to a public dashboard that is cross-linked to the corresponding source code repositories.

**Use Case: Bioconductor**

**Author of Use Case**: Guergana Savova, BCH/HMS; Martin Morgan, Roswell Park Comprehensive Cancer Center

**Short History**: Bioconductor provides tools for the analysis and comprehension of high-throughput genomic data. Bioconductor uses the R statistical programming language, and is open source and open development. It has two releases each year that follow the semi-annual releases of R. The project started in 2001 at the Dana Farber Cancer Institute. It matured at the Fred Hutchinson Cancer Research Center between 2004 and 2015. The core team has been at Roswell Park Cancer Institute since then.[1-3]

Bioconductor supports many types of high-throughput sequencing data (including DNA, RNA, chromatin immunoprecipitation, Hi-C, methylomes and ribosome profiling) and associated annotation resources; contains mature facilities for microarray analysis; and covers proteomic, metabolomic, flow cytometry, quantitative imaging, cheminformatic and other high-throughput data.[4]

**Governance**: A Technical Advisory Board of key participants meets monthly to support the Bioconductor mission by developing strategies to ensure long-term technical suitability of core infrastructure, and to identify and enable funding strategies for long-term viability. A Scientific Advisory Board including external experts provides annual guidance and accountability.

**Documentation**: Bioconductor documentation comes at three levels: workflows that document complete analyses spanning multiple tools; package vignettes that provide a narrative of the intended uses of a particular package, including detailed executable code examples; and function manual pages with precise descriptions of all inputs and outputs together with working examples. In many cases, users ultimately become developers, making their own algorithms and approaches available to others.

**Code quality**: All Bioconductor packages are built nightly on Windows, macOS, and Linux platforms. Packages must build successfully before propagating to public user-facing repositories. A successful build requires that the package pass a suite of tests defined by R. The tests ensure overall package integrity (e.g., a DESCRIPTION file describing author, title, abstract, software dependencies, license, etc.; correct package structure and language syntax, etc.), functionality (e.g., correct example, vignette, and test code evaluation) and integration with current versions of package dependencies.

**Support**: Users navigate Bioconductor through views https://bioconductor.org/packages/release/BiocViews.html#___Software (a directed acyclic graph of terms from a controlled vocabulary) linking to individual 'landing pages' (e.g., DESeq2 https://bioconductor.org/packages/release/bioc/html/DESeq2.html) providing an overview of the package, installation instructions, usage statistics, etc. A user-oriented StackOverflow- style support site https://support.bioconductor.org is available and active (100's of visitors per hour; generally fast response, e.g., < 1 day, by package authors or experts). All packages include maintainer email addresses; many include links to, e.g., GitHub issues. Developer support is provided by the bioc-devel email list-serve https://stat.ethz.ch/mailman/listinfo/bioc-devel. A hybrid community slack https://bioc-community.herokuapp.com/ is proving to be an effective tool for fostering experienced user / developer collaboration.

**Ecosystem collaboration**: There are 1649 packages from the core group and the international community. Packages have been contributed by more than 1000 maintainers. There are >29,000 mentions of "Bioconductor" in the scientific literature, and 500,000 unique IP addresses in 2018.

**Legal**: Almost all Bioconductor packages are licensed in a way that allows use by any entity without permission; a small number (<10?) specify 'academic only' use. Figure 1 lists the licenses associated with R packages in Bioconductor. The main license groups are:
(a)      Artistic license v2 (https://opensource.org/licenses/Artistic-2.0 ) which appears commercially friendly
(b)      GPL which requires that distribution of any modified code must make the source available
(c)      MIT, BSD, CC fall in the group of lax permissive licenses
(d)      Other

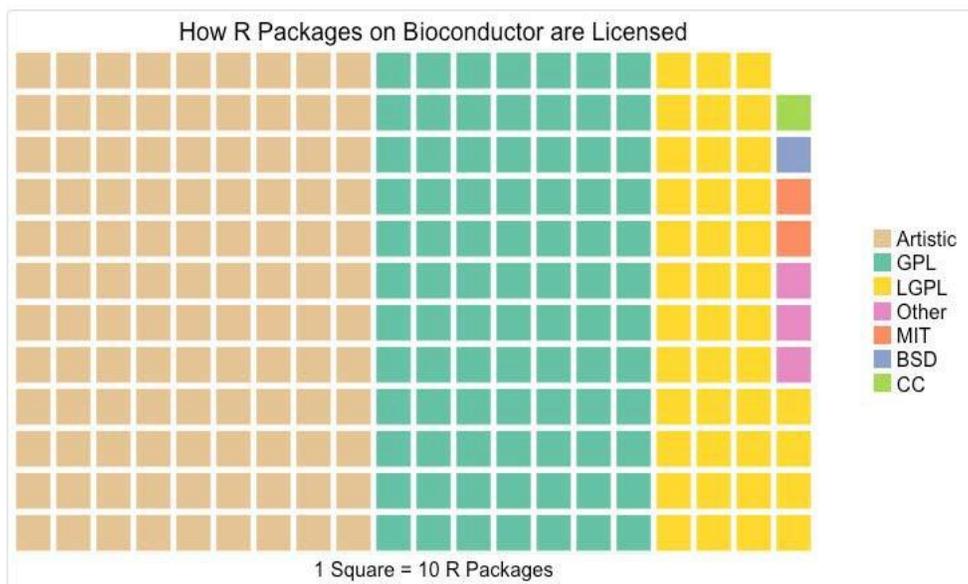

*Figure 1: Bioconductor packages and their licenses (source: http://seankross.com/2016/08/02/How-R-Packages-are-Licensed.html )*

**Finance**: Initial support was from Institutional funding. NIH / NHGRI support began in 2003 with BISTI grant R33HG002708, and has continued via P41 / U41HG004059 since 2006. NCI / ITCR funding began in 2014 with U24CA180996. Additional grants have been awarded to key stakeholders in the Bioconductor community, with effort directed in part to core project activities; funding sources have included the NSF, European Union, NCI ITCR, NHGRI, and the Silicon Valley Foundation. Details are available in annual reports[4]; these funding sources have not been pursued in a coordinated fashion, but in response to needs and interests of the community. Many individual packages represent the product of federally funded researchers outside the core team; surprisingly, funding sources are almost never acknowledged in contributed packages.

**Marketing**: Bioconductor primarily relies on provision of current and relevant software resources as the primary marketing approach. Bioconductor has a consistent web presence and annual conferences. Social media (e.g., twitter) and video sharing (YouTube) channels exist, but are not a primary means of engaging new users.

**Dependency hygiene**: Most Bioconductor packages have dependencies on other R packages. These packages must come from either Bioconductor or CRAN, repositories with conservative standards and commitment to long-term availability. R packages can have system dependencies, typically operating system libraries. Rarely, packages depend on third-party software, or software written in other languages, e.g., python. Bioconductor packages generally have more R package dependencies than non-Bioconductor packages. This deeper dependency graph arises in part from emphasis on re-use of robust and interoperable core- developed infrastructure. Some contributed packages try to capture A-to-Z 'workflows'; these tend to have a large number of dependencies, are fragile and, perhaps ironically, have limited audiences.

**References**
1. https://www.bioconductor.org/
2. https://en.wikipedia.org/wiki/Bioconductor
3. Annual reports at https://bioconductor.org/about/annual-reports
4. Huber, W. et al. Orchestrating high-throughput genomic analysis with Bioconductor. Nat. Methods 12, 115–121 (2015).

**Use Case: Cytoscape**

**Author of Use Case:** Juli Klemm

**Short History of the Tool:** Cytoscape was originally created at the Institute of Systems Biology in Seattle in 2002. Now, it is developed by an international consortium of open source developers. Cytoscape was initially made public in July, 2002 (v0.8); the second release (v0.9) was in November, 2002, and v1.0 was released in March 2003. Version 1.1.1 is the last stable release for the 1.0 series. Version 2.0 was initially released in 2004; Cytoscape 2.83, the final 2.xx version, was released in May 2012. Version 3.0 was released Feb 1, 2013, and the latest version, 3.7.0, was released in October 2018.

**Governance:** Cytoscape is architected as Core software (maintained by the Core team) augmented by Cytoscape Apps (contributed by community members. The Cytoscape apps are mechanisms for defining and delivering novel calculations and workflows and are available through the Cytoscape App Store, http://apps.cytoscape.org/. The Core Team are developers who have GitHub commit access to the Core software.
Cytoscape 3.0 has a clearly defined, simplified API. API jars are strictly separate from the implementation jars. The API is versioned using the Semantic Versioning standard. This means that the API won't change throughout 3.x, so an app designed to work with an early version of 3.x will be guaranteed to work up to version 4.0 of Cytoscape. Cytoscape maintains an explicit backwards compatibility contract found in each class in the public API so that both core developers and app writers will understand how a class might change.
https://github.com/cytoscape/cytoscape/wiki/Cytoscape-3.0-App-Development

**Documentation:** Cytoscape has extensive user and developer documentation. The developer resources (https://cytoscape.org/documentation_developers.html) include an issue tracker, Nexus repository, nightly builds, and code metrics. In addition, there is a Cystoscape "App Ladder" that teaches essential skills for developing Cytoscape Apps,
https://github.com/cytoscape/cytoscape/wiki/Cytoscape-App-Ladder.
User resources include basic and advanced tutorials, a YouTube channel, a blog of published figures, as well as several presentations all available
athttps://cytoscape.org/documentation_users.html

**Code Quality:** Cytoscape core developers use Jenkins, and open source automation tool to support Continuous Integration. It is used to build and test software projects continuously to make it easier for developers to integrate changes to a project. http://code.cytoscape.org/jenkins/
Regarding apps, contributors are strongly encouraged to thoroughly test apps before releasing them. However, the Cytoscape team does not independently review them.

**Support:** Cytoscape Help Desk: https://groups.google.com/forum/#!forum/cytoscape-helpdesk

**Ecosystem Collaboration:** Cytoscape has an active and diverse community of developers and users. Communication happens mostly via mailing lists and social media:
o   The App Developer Mailing List is for asking questions about app development.
o   The Announcement Mailing List is used for announcements from the core development team

- to the rest of the community.
  - The Help Desk is where users can ask questions on usage.
  - Cytoscape Twitter.
  - The Cytoscape Publications Tumblr highlights published research using Cytoscape and published plugins.

**Security Legal:** Cytoscape is available as a platform-independent open-source Java application, released under the terms of the LGPL. By downloading Cytoscape, you agree that you have read the license agreement that follows and agree to its terms. If you don't agree, do not download Cytoscape.

**Finance:** Funding for continued development and maintenance of Cytoscape is provided by the U.S. National Institute of General Medical Sciences (NIGMS) under award number R01 GM070743. Cytoscape user support, education and new initiatives are supported by the National Resource for Network Biology (NRNB) under award number P41 GM103504.
Cytoscape consortium http://169.228.38.215
The Cytoscape Consortium holds the intellectual property rights of the following software:
- Cytoscape (LGPL) – www.cytoscape.org
- Cytoscape.js (LGPL) – http://cytoscape.github.io/cytoscape.js/NeXO (BSD or similar, in progress) – http://nexontology.org/
- NDEx (BSD or similar, in progress) – www.ndexbio.org

**Marketing:** Cytoscape is marketed through Twitter, Tumblr, Vimeo, and Pinterest as well as mailing lists and forums for specialized audiences.

**Dependency Hygiene:** Note that some Cytoscape functions rely on code deployed as services available on web servers. Generally, such services are callable by Cytoscape or directly by non-Cytoscape clients (e.g., Python) in the larger bioinformatics community. Some services are provided by other organizations (e.g., PSICQUIC for importing public networks), while others are provided by Cytoscape developers (e.g., Diffusion) and are located in or rely on other GitHub repositories. Here is a list of known external repositories containing services called by Cytoscape and maintained by Cytoscape core developers:
- CXMate - adapters that simplify service writing https://github.com/cxmate/cxmate
- Diffusion - called by Diffusion core app https://github.com/idekerlab/heat-diffusion

Each repo contains information on how to build and deploy the service.

**Use Case: Globus**

https://www.globus.org/

**Author of Use Case**: Silverstein JC (U Pitt), Davis M (U Pitt), Espino J (U Pitt), Ian Foster (U of Chicago), Gilbertson J (U Pitt), Raumann B (U of Chicago), Taylor D (U Pitt), Zelnis J (U Pitt) and Becich MJ (U Pitt).

**Short History of Tool**: The first public Globus product was the Globus Toolkit: software for enabling distributed computing. The Globus Toolkit was conceived to remove obstacles that prevent seamless collaboration by allowing users to access remote resources while simultaneously preserving local control over who can use resources and when. Foster et. al. continued to develop the Globus Toolkit for fifteen years, trying out several sustainability strategies. The initial releases in 1998 and 1999 were open source, and from the beginning, there were contributors who helped make it useful to researchers. In 2003, an international research consortium, the Globus Alliance,[1] was formed with the hopes of encouraging broader research and academic support. In 2004, the company Univa[2] was launched with the hopes of raising money by offering support services for enterprises. In 2005, an industry organization, the Globus Consortium,[3] with members including IBM, Sun, Hewlett-Packard, and Intel, was formed. Despite these efforts, the pressure to sustain the Globus mission remained.

In 2009, a new model was adopted with three key elements: a tighter focus on data management for researchers, a "freemium" support model,[4] and cloud-based Software-as-a-Service (SaaS) delivery.[5] In 2010 the new service was launched, which is now known simply as Globus. SaaS delivery allowed features needed by researchers that the software toolkit could not provide, such as 24/7 availability, little or no installation or configuration, and simple, intuitive web interfaces, and control of (and responsibility for) the end-to-end user experience.[6,7] Globus is operated as a non-profit service within the University of Chicago, in order to remain focused on addressing the unique needs of the non-profit research community.

Under the current freemium model, Globus services are free for researchers while institutions pay an annual subscription fee for access to premium features. Premium features are designed to offer value to research computing directors and, by extension, to the research community they serve. The first subscription was signed in 2013, and, in 2019, the 100th subscriber was signed. Nonetheless, Globus subscribers still only represent a small fraction of those institutions that use Globus worldwide. The freemium approach is essential to the goal of sustaining Globus as a non-profit service for the non-profit research community. By thus operating Globus as a low-cost, "freemium" cloud service, the longevity is ensured of what has become a vital infrastructure component for many resource owners, HPC facilities, and researchers.

The research community is characterized by many small teams with specialized needs that require custom applications. Research application and service developers establish a growing community that is clearly important to the future of research. These developers need a platform that enables large-scale and automated data management. The Globus service is built on an emerging Platform-as-a-Service (PaaS)[8]: a collection of services with REST APIs,[9] designed for research application developers, and accessible via a software development kit (SDK) and a command-line interface (CLI). Globus SaaS itself uses the public Globus PaaS for building and operating the primary Globus service. Substantial investment is planned in expanding the platform's functionality so that developers can build richer solutions for their research communities.[10]

**Governance:** The Globus project is governed by its PI, Ian Foster, and the University of Chicago. The Globus SaaS is operated out of the University of Chicago as a service core to the non-profit research community. Technical, business, and management decisions are the responsibility of Globus staff. Institutional subscribers via subscriber meetings and consultations, and community users via mail lists, webinars and on-site tutorials, provide input on feature prioritization, product roadmap, subscription pricing, and sustainability models.

**Documentation:** Globus provides fully developed documentation for end-users, developers and administrators. This documentation adequately covers a wide range of topics including guides for installation, FAQs, API usage, tutorials, etc.[11] Each software release is clearly versioned with release notes and change history documented.

**Code quality**: Globus has a professionally maintained and managed code base that is developed with best practices such as version control and software management in a source code repository (GitHub), extensive documentation and code comments, linting, distribution of knowledge across team members, and incremental/iterative software development. Code is reviewed by at least one other software engineer before release to production. Globus employs a continuous integration environment, automated tests, and documented, standardized human QA testing to ensure code quality. Globus developers follow secure development practices, including OWASP recommendations to prevent web application security risks.

**Support**: Globus provides several support options from online self-help tools, listserv groups to a ticketing submission system with a responsive, dedicated support team.[12] Globus subscribers are guaranteed a response time of one business day to support tickets.[13]

**Ecosystem collaboration:** Globus has a large user base of over 200,000 registered users and 140 institutional subscribers. Globus is widely used across the physical, social, and life sciences, including at most leading US universities and many sites overseas. Globus moves thousands of terabytes a day of research data, and there are 2,000 active multi-user storage systems (e.g., campus storage systems) and over 20,000 active single user storage systems (e.g., laptops) accessible via Globus. A strong community is supported by multiple channels of user engagement. Daily interaction between users and the Globus team occurs through the robust ticketing system and active list serv groups. Numerous webinars and on-site, two-day workshops occur every year at research institutions across the US.[14] The multi-day annual user conference provides a forum for users and subscribers to connect with the Globus team as well as with each other. The conference includes a half day meeting for subscribers to provide input into feature priorities, overall product roadmap, subscription pricing, and sustainability models.[15] The Globus customer engagement team facilitates application of current Globus services by ensuring that institutions provide the maximum value of their subscription to their researchers. The Globus professional services team engages in grant-funded collaborations with research teams to apply, extend, develop, or innovate Globus services in the context of science drivers.

**Security:** Globus has maintained a strong security model for many years, using highly standards-based components and protocols that address message protection, authentication, delegation and authorization for distributed infrastructures. Globus has adopted new technologies and procedures to ensure the continued security of data exchange, for example the evolution from WS-security-compliant message-level and credentials security using X.509 to the current model of OAuth, OpenSSL and other new standards and the increased security posture through adoption of more rigorous security controls.

Globus services leverage federated login and allow user authentication using one of the many supported identity providers (e.g., institutional identities, eRA Commons, ORCID, Google). Since Globus acts as an identity broker and uses federated login, institutional credentials are never seen by Globus. Globus Auth, the identity and access management service in Globus, is based on OAuth 2 and is used to secure all Globus services and used for integration with third-party applications. This provides an advanced, user consent based delegated authorization model that allow applications and services to act on behalf of users and other services.

Globus uses a "data channel" for moving data between two endpoints. This data channel is established directly between the source and destination endpoints and cannot be accessed by the Globus service, only by the servers running on the endpoints. Transfers can be encrypted using OpenSSL libraries installed at the endpoint and TLS 1.2. Transfer of designated restricted data are always encrypted. In addition to the data channel, Globus uses a "control channel" to communicate with the source and destination endpoints for a transfer. The control channel is encrypted with TLS 1.2 or higher.

In 2017, the Globus introduced a high assurance tier that provides additional security controls to meet the higher authentication and authorization standards required for access to restricted data, such as Protected Health Information, Personally Identifiable Information, and Controlled Unclassified Information. Users must authenticate with specific identities as determined by the policy set by administrators at the institution to obtain access. In additions, users must re-authenticate in each new application session with the required identity and each authentication lasts for a specific period of time after which the user must re-authenticate. A detailed audit trail is generated that allows reconstruction of data access and user activities.[16]

Globus has undergone security reviews by various external bodies, including source code reviews by the NSF Trusted CI center and HIPAA risk and gap analyses by third party auditors.

**Legal:** The Globus legal approach to software distribution has evolved with the Globus technology. Currently Globus uses a variety of licensing models. Select software components operated as a service and client-side software, such as the Globus Command Line Interface, the Globus Software Development Kit, and some version of Globus Connect (the software that enables storage systems to be accessed via the Globus service) are made available under the Apache 2.0 license. Some versions of Globus Connect are licensed under the Globus Community License, under which subscribers can access source code for the purposes of code review and contribution.[17] Globus does not license the remaining software components it operates as a service.

Globus holds no patents on Globus technology. Globus has proposed that the University of Chicago enter into a software code escrow agreement with its federal funding agencies that would provide all Globus source code and related materials be transferred to the research community if the Globus team is no longer willing or able to provide the Globus service to the non-profit research community. Globus would transfer all the relevant architecture documents and artifacts (including source code, design documents, test plans, files, working plans, database designs etc. when applicable) for all information systems and provide assistance and open communications until the transfer were complete.

**Finance:** To enable long-term sustainability and alignment with science needs, Globus is provided to the community via a freemium model. Data transfer is free; other services (e.g., data sharing, usage management services for research computing administrators) are provided under an annual subscription by the institution. This subscription model is designed to allow research institutions to engage with Globus as subscribers, and then make premium Globus services freely available to their researchers, educators, and students. Furthermore, the basic Globus data transfer capability is made freely available to researchers and educators in the non-profit sector. Thus, any collaborator of researchers at a subscribing institution can access data that those researchers make available via Globus. Globus focuses non-profit research use; however, commercial use is allowed.[13]

Subscriptions are a flat annual fee that allow unlimited use. The fee is based on the institution's level of research activity, in order to ensure equitable pricing. Hence, large research institutions pay more than a regional college with a limited research program. In the US, pricing is determined by the institution's Carnegie Classification (carnegieclassifications.iu.edu); for non-US institutions, an attempt is made to find a suitable proxy by looking at the number of researchers served and the research budget. Pricing is higher for commercial organizations and subscription levels are structured differently for commercial use (most notably, commercial subscriptions are not unlimited use).

Although under the freemium model an increasing portion of ongoing funding has shifted from federal grants to institutional subscriptions, federal funding is still important for developing new capabilities—which are then sustained via subscriptions.

**Marketing:** Globus attributes their brand recognition primarily to sustained, programmatic outreach via multiple channels, with a set of consistent messages, and a long history of high quality. In addition, Globus brand recognition is strengthened by a memorable, uncomplicated name and a simple logo. Time and resources have been invested into a well-designed public website that is optimized for search.
A variety of communication methods keep Globus visible and the user community updated:
o Speaking engagements and participation in workshops, working groups, and conferences
o Presence on social media (Twitter and LinkedIn, primarily
o Presence in news media via occasional press releases and article postings
o Blog with frequent (ideally one to two per month) announcements and articles
o Email announcements to opt-in marketing database that include monthly newsletters and periodic announcements for new blogs, news releases and events
o Email announcements to user and admin email lists maintained
o Engagement with partners (e.g. via joint webinars, event sponsorship, etc.)
o Workshops and user conferences, for example the annual Globus users conference and on-site tutorials and workshops.

**Dependency Hygiene:** Most to all software has some dependencies on other software or libraries, Globus is no exception. However, the transition to a SaaS model has greatly reduced the software dependencies.  Only Python libraries for Globus Connect Server and OpenSSL and SSH for Globus Connect Personal are included in shipped software.  These dependencies are generally updated with each software release. During installation, the user is prompted to install dependencies that are not included in Globus software packages. Updates of these dependencies is the responsibility of the storage administrator, who can get the list of these dependencies by querying the Globus software packages.

**Use Case: i2b2 tranSMART**

**Author of Use Case:** Jonathan Silverstein (U Pitt), Michael J. Becich (U Pitt), Keith Elliston (tranSMART Foundation & Axiomedix, Inc.), and Ye Ye (U Pitt)

**Short History of Tool** The i2b2 project was established in 2004 as an NIH-funded National Center for Biomedical Computing at Brigham and Women's Hospital and was later based at Partners HealthCare System. The i2b2 Foundation was formed in 2016, to provide oversight and governance for the i2b2 project. The i2b2 platform develops a web-enabled warehouse server and software tools, enabling researchers to retrieve patient cohorts and obtain project-specific databases (integration of medical record and clinical research data) while preserving patient privacy.[1] This platform also provides a set of web services to load additional data, such as applying natural language processing service to determine patients' smoking history from discharge summaries.[2] Over 250 hospitals and research centers have adopted the i2b2 platform for sharing, integration, standardization, and analysis of heterogeneous data from healthcare and research. i2b2 facilitates three different data integration approaches: one based on ontology, one based on the tranSMART engine, and the third based on CouchDB.[3]

The tranSMART Foundation was established in 2013 as a public-private partnership between scientists in the United States and the European Union. Founding partners include the University of Michigan, the Pistoia Alliance and Imperial College London. The initial version of tranSMART's data management system was developed in 2009 by scientists at Johnson & Johnson and Recombinant Data Corporation, and mainly used by pharmaceutical researchers. TranSMART has being developed to support one informatics-based analysis and pre-competitive data-sharing platform for clinical and translational research,[4] enabling hypothesis generation and validation using integrated data.[5] The development currently focuses on TraIT and eTRIKS, which are based on the tranSMART platform, supported by major European translational research initiatives.

In May 2017, the i2b2 Foundation (Informatics for Integrating Biology and the Bedside) and the tranSMART Foundation merged to create a single, unified organization, i2b2 tranSMART Foundation to advance the field of precision medicine. In 2019, the i2b2 tranSMART Foundation worked with Partners Healthcare to release i2b2 under the weakly permissive MPL2 open source license (previously i2b2 was distributed under the BWH License, with was not Open Source Definition compliant).

**Governance:** The governance structure of the i2b2 tranSMART Foundation is defined in their Bylaws available publicly https://drive.google.com/file/d/0B8lizkKDeaKhTUF3QmNTTFk0ZnM/view. The governance structure makes the foundation not rely on any single person. High-level leaders include board of directors, executive committee, governance/nominations committee, finance/audit committee, and technology committee. The membership program includes stakeholders based on merit and contribution, and this group is actively engaged in operations, fundraising, software development, and the nomination and election of new board members. In addition, members participate in working groups, such as ETL working group, ontology working group, and user interfaces working group.

Each of the software projects organized by the i2b2 tranSMART Foundation are governed by Project Management Committees (PMCs) that are modeled after the "Apache Way." These PMCs provide coordination of development, establish standards for testing, and manage the release process of the platforms.

The i2b2 tranSMART foundation provides many ways to involve new users and contributions, including free training class, working groups open to anyone, sign-up mailing list, sponsorship programs, user platforms, and vendor sign up link. Monthly webinars have been organizing since March 20, 2018, (and since 2014 under the predecessor tranSMART Foundation) during each of which a community member presents an update on the foundation and a preview of upcoming events. These recorded webinars are publicly available online through YouTube https://www.youtube.com/channel/UC3hy0Az4VYs6TZgN79-w8TA.

**Documentation:** The i2b2 community provides updated documentation through a Wiki page https://community.i2b2.org/wiki/pages/viewpage.action?pageId=342684. The documentation is organized for different interest groups. For beginners, documentation includes installation guide, upgrade guide, and tutorials. For developers, documentation includes server-side messaging, server architecture, server-side design, web client design, and release notes. For end users, documentation includes web client help and workbench user guide. The i2b2 documentation lists changes of each release and bug fixes. For each bug fix, i2b2 documentation provides a bug ID, a one sentence summary, and details about the content and reasons of changes and affected versions. Commit messages in the i2b2 GitHub https://github.com/i2b2 are simpler, listing changes in codes. The i2b2 provides human-focused release notes for every release.

The tranSMART community provides documentation on its current version 16.3. The documentation includes instructions on different functions: browse, analyze, summary statistics, advanced analyses and visualizations, export results, sample explorer, gene signature wizard, and genome-wide association study tool. The tranSMART GitHub website https://github.com/transmart includes commit messages for each change of codes.

**Code Quality:** Both i2b2 and tranSMART have extensive automated and manual testing as a part of their well define release process. Example test plans for i2b2 include: https://community.i2b2.org/wiki/display/HOM/HOM+Home?preview=%2F336164%2F336376%2FUETL+Test+Plan.docx

**Support:** The i2b2 foundation has established an i2b2 Bug Tracker for contributors to receive suggestions for improvements. https://community.i2b2.org/jira/secure/Dashboard.jspa. Outside users can sign an account to review the records of historical and current issues and solutions. The i2b2 Bug Tracker system flags important issues as *ToDo*, *In Progress*, *Done*, and *Other Issues*. It also allows each user to track his/her specific issues. In addition, a publicly available Google Forum is organized to assist installation https://groups.google.com/forum/#!forum/i2b2-install-help.

TranSMART has a platform for developers and testers through tranSMART Wiki https://wiki.transmartfoundation.org/display/transmartwiki/tranSMART+Project+wiki+Home, which is not currently accessible.

**Ecosystem Collaboration:** The i2b2 tranSMART Foundation provides a platform for contributions of new tools. Guidelines of submitting new contributions are available through https://transmartfoundation.org/transmart-platform-code-contributions/.

In addition, the i2b2 platform provides information of finished and ongoing community projects through https://community.i2b2.org/wiki/display/i2b2/i2b2+Community+Projects. For example, the i2b2 FHIR Cell project allows i2b2 core to communicate with SMART cells using the Fast Healthcare Interoperability Resources (FHIR).

**Security:** In the i2b2 platform, securable remote access is provided through well-defined messages in web services. Anonymous patients' aggregate counts obfuscated by adding small random numbers are returned in queries in order to protect patients' privacy.[1] Moreover, the i2b2 PM Cell Security Enhancements project consists of modifications to the i2b2 core project management cell in order to increase the security of the authentication information https://www.i2b2.org/software/contributed.html. The security enhancement is becoming the focus of the next core release.

The tranSMART platform also has a mechanism to manage security for studies. The protect study operation needs to receive an approval of administrator. After a user loads study data into a database server, tranSMART administrator will assess whether a study is a secure object or not. If a study is deployed on multiple servers, then the authorization must be conducted on each server separately https://transmart.support.axiomedix.com/hc/en-us/articles/360005847334-Managing-Security-for-Studies.

**Legal:** The i2b2 and tranSMART are free open source software https://transmartfoundation.org/legallicensing/
The i2b2 software is licensed through the Mozilla Public License (MPL) version 2.0 https://www.mozilla.org/en-US/MPL/2.0/ under the terms of the Healthcare Disclaimer addendum https://community.i2b2.org/wiki/display/webclient/The+i2b2+MPL+2.0+License+with+Healthcare+Disclaimer+Addendum
The tranSMART software is now made available under the terms of the version 3 of the GNU General Public License (GPL v3) https://www.gnu.org/licenses/gpl-3.0.html

**Finance:** To help financially support the i2b2 tranSMART operations, the Foundation offers four sponsorship programs, including contributing sponsors (medical and academic research centers and small vendors), corporate sponsors (large corporate supported), sustaining sponsors (organizations providing on-going staff and financial support), event sponsors (vendors who offer product or services). https://transmartfoundation.org/contributing-sponsors-program/
- Contributing sponsors: Michigan Medicine, University of Michigan, U of Kansas Medical Center, ITTM (Information Technology for Translational Medicine), Beth Israel Deaconess Medical Center, University Medical Center Göttingen, Prognosis Data, InterSystems, Boston Children's Hospital Computational Health Informatics Program, Wake Forest Clinical and Translational Science Institute
- Corporate sponsors: Takeda
- Sustaining sponsors: Department of Biomedical Informatics, Harvard Medical School, Partners Healthcare, Harvard Catalyst, Axiomedix

- Event sponsors: Department of Biomedical informatics, Harvard University, Harvard Catalyst, Persistent, Essex, Axiomedix

Complementary with the nonprofit sustainability model through the tranSMART and the successor i2b2-tranSMART Foundation efforts, the team that founded the tranSMART and i2b2 tranSMART Foundations, Keith Elliston and colleagues, started Axiomedix in 2018 specifically to offer a four-part commercial support mechanism to sustainably support open source software projects, particularly after their grant funding cycles have ended:

First, Axiomedix offers a commercial grade software publishing and support model. For this, Axiomedix works with developers to create a supportable, tested and validated version of the open-source platform, running on a commercial grade and updated technology stack. So far, Axiomedix supports i2b2 and tranSMART, and is working to support a number of other platforms (Arvados, CWL, openBEL and others). In this model, Axiomedix pays a royalty on all sales to the OSS project (if applicable and possible) and pays a royalty to the developers in return for their work to develop and support the commercial version, providing tier III and IV support through the Axiomedix support portal (support.axiomedix.com).

Second, Axiomedix offers full-service solutions, which include installation, configuration, data loading, curation, training and more. Axiomedix works directly with developers to provide these services, as high-value contract work.

Third, Axiomedix has developed the Axiomedix Expert Network, which is a network of core developers that can perform part-time, contract work with commercial customers on an hourly basis. This enables an efficient development of new capabilities for the platforms, and a direct integration of these enhancements into the core open-source codebase. For all of these efforts, Axiomedix developed close relationships with the core developers of the open-source platforms, which can provide them with sustainable income streams even after the grant funding for these projects has ended.

Fourth, Axiomedix has brought together highly skilled technologists, open source technologies and subject matter experts, to develop new software products that enable core aspects of precision medicine. These projects can be funded by grants, sponsors or investors, and can result in new open source products or commercially developed products. This approach enables the development of new innovations on the core capabilities of the scientific open source ecosystem.

**Marketing:** The i2b2 tranSMART foundation has a logo, website, and its own trademarks. Official announcements are available through https://transmartfoundation.org/news/. The foundation has set up many workgroup mail lists. The tranSMART platform has additional system administrator and developer mail lists. Moreover, the foundation has established many communication channels to announce and highlight interesting events and topics:
Twitter: https://twitter.com/i2b2tranSMART
LinkedIn: https://www.linkedin.com/groups/?home=&gid=4218734
Facebook: https://www.facebook.com/i2b2tranSMARTfoundation/
Google forum: https://groups.google.com/forum/#!forum/transmart-discuss
YouTube channel: https://www.youtube.com/channel/UC3hy0Az4VYs6TZgN79-w8TA

**Dependency Hygiene:** Both the i2b2 workbench developer's guide and i2b2 installation guide provide information about dependencies. In addition, the i2b2 developers create three Docker contains "i2b2-web," "i2b2-wildfly," and "i2b2-pg" to encapsulate the core functionalities and facilitate the configuration of numerous dependencies in i2b2 components.[6]

   The TranSMART ETL guide https://www.etriks.org/wp-content/uploads/2015/12/etl-getting-started-4.pdf lists dependencies, such as pentaho data-integration software suite version 4.4.0., Java Runtime Environment (JRE), SSH client for secure connection, postgres client to database connection. Each release is analyzed for dependencies, and to ensure that there is a definitive list of components within the system. TranSMART Docker is available at https://github.com/hms-dbmi/transmart-docker

## Use Case: ITK

https://itk.org

**Author of Use Case:** Sarachan BD (GE Research), Miller JV (GE Research), Fedorov A (BWH).

**Short History of Tool:** In 1999 the National Library of Medicine (NLM) and NIH awarded a three-year contract to develop an open-source image registration and segmentation toolkit, which eventually came to be known as the Insight Toolkit (ITK). The images in question are typically from medical images such as CT or MRI scanners. Registration refers to aligning the or developing correspondences between image data. Segmentation refers to identifying and classifying data in the images, such as nodules in lung. An original goal of ITK was to process data from the Visible Human project which included CT, MRI, and cryosections. Since then, it has been applied successfully for the analysis of a broad range of imaging modalities, including non-clinical images, such as electron microscopy. ITK is implemented in C++ and provides a class library for developing software applications involving image registration and segmentation.

The ITK consortium members included three companies: GE Research, Kitware, Inc., and MathSoft (now called Insightful), and three academic members: University of North Carolina, University of Tennessee, and University of Pennsylvania. Each organization had its own PI. The GE PI was Bill Lorensen, a GE Coolidge Fellow, which is the highest honor awarded to GE Research employees. The overall program Project Manager was Dr. Terry Yoo from NLM. Other contributing organizations include Brigham & Women's Hospital, Columbia University, and University of Pittsburgh. There were many NIH companion grants and contracts that further developed ITK.

ITK has been used in a number of commercial products from GE and likely other companies, ITK has also been used internally to GE to build important internal applications, such as for industrial inspection, even beyond the healthcare industry. GE maintains its own extension to ITK consisting of algorithms that are either proprietary or not generalized enough for community use.

ITK has also been instrumental as a component to numerous end-user medical image computing research tools. Some of the most prominent examples of such software packages include 3D Slicer (https://slicer.org) and MITK (http://mitk.org/).

Although ITK is a C++ library, there are several packages that provide Python packaging (i.e., SimpleITK and itk-python) further increasing usability of the software and integration with the other widely popular Python tools.

**Who or what is ITK's competition?**

There are commercial services that provide clinically-approved image diagnostics and commercial software and services for developing purpose-specific image analysis algorithms (e.g. Definiens). We are not aware of an open source competitor to ITK having its breadth of capabilities and flexibility specifically for radiology. Packages such as ImageJ are popular for microscopy image viewing and analysis. ITK has been used a component of several other open source software.

**Governance:** A non-profit ITK consortium has been formed that manages ITK licensing, owns the copyright, and acts as a governing body. Code changes are controlled by strictly limiting who has access to commit changes to the repositories. There is a code review process established for submissions. An online ITK journal documents changes and acts as a forum for proposed ITK changes. ITK development is maintained fully via GitHub, which hosts the source code, release packages, and bug tracker (see https://github.com/InsightSoftwareConsortium/ITK). Contribution guidelines are documented, and there are numerous contributions from the community to its content.

**Documentation:** There is much inline documentation as well as a user's guide that has been published as a book. Coding example are automatically built as part of nightly tests. There are strictly- enforced coding conventions such as consistent naming rules.

**Code quality:** ITK had automated nightly builds and tests as far back as 1999, being an early adopter of this software-engineering best practice, before the widespread adoption of continuous integration and GitHub

**Support:** ITK has mailing lists and over its long history has had dedicated volunteers who would give detailed help to users even including example code. ITK has its own Discourse forum for discussions (https://discourse.itk.org) and mutual help among users. Very importantly, NIH has continued to provide maintenance contracts for bug fixes, incremental improvements, and a moderate level of user support (i.e. brief answers to questions). This maintenance has typically been performed by Kitware, providing continuity and expertise. Kitware also offers commercial ITK support for a fee.

**Ecosystem collaboration:** There have been many contributions to ITK over time, including community members not specifically funded for contributing to ITK. As of this writing, there have been over 200 developers contributing to the source code of ITK, according to the automatically collected statistics by GitHub, and available publicly at https://github.com/InsightSoftwareConsortium/ITK/graphs/contributors.

**Security:** Security has not been a primary consideration for ITK, which is typically embedded within broader systems. That being said, ITK is likely not particularly vulnerable to cyber-threats. Being implemented in a low-level language (C++) some types of memory hacks may be possible.

**Legal:** Initially ITK was licensed under BSD. This was later changed to Apache which requires contributors to abide by the open source requirements.  The Debian operating system distributes ITK and performs scans to ensure there are no embedded third-party packages having license restrictions. Initially ITK permitted the inclusion of patented software, but after the administrative of these special cases became burdensome, patented materials were no longer permitted for inclusion.

**Finance:** As mentioned above, ITK is free to users, NIH has provided continual funding for maintenance, and commercial-grade support can be purchased.

**Marketing:** Initial visibility for ITK was via promotion at medical imaging conferences. Over time, the use of ITK for medical image analysis became pervasive.

**Dependency Hygiene:** ITK has managed its dependencies using a name-mangling scheme that would prevent third- party packages from having name conflicts. This would be somewhat laborious for the ITK developers but resulted in ITK being very easy for users to build. ITK does not have a built-in package manager such as pip for Python.

References

1. https://itk.org
2. https://en.wikipedia.org/wiki/Insight_Segmentation_and_Registration_Toolkit
3. https://www.kitware.com/
4. https://www.definiens.com/

34 | P a g e

**Use Case: Linux**

**Author of Use Case**: Guergana Savova, BCH/HMS

**Short History**: Linux is the world's largest and most pervasive open source software project in the history of computing. The Linux kernel which was released by Linus Torvalds in 1991 is the largest component of the Linux operating system. Linux has become the world's most dominant operating system with adoption in various sectors including finance, government, and education. "it is also the operating system of choice to support cutting-edge technologies such as the Internet of Things, cloud computing, and big data."[1]

**Governance**: "Linux is the premier example of open source sustainability and success. The non-profit Linux Foundation founded in 2000 provides a neutral home where Linux kernel development can be protected and supported for years to come:
- Linux Foundation fellowships sponsor the work of the Linux creator Linus Torvalds and lead maintainer Greg Kroah-Hartman
- Linux Foundation IT operations run the systems behind Linux kernel development on kernel.org
- Linux Foundation Training offers free and paid training courses and Linux certifications
- Linux Foundation Events organize gatherings where kernel developers can collaborate"[1]

The Linux Foundation Board of Directors is comprised of 22 senior leaders from across the IT industry. Board members represent Linux Foundation members and the Linux developer community and set the strategic direction for the organization.

**Documentation**: The Linux kernel provided user's and administrator's guides available at https://www.kernel.org/doc/html/latest/ Code is available at https://github.com/torvalds/linux

**Support**: Support is provided through the LF JIRA at https://support.linuxfoundation.org/

**Ecosystem collaboration**: "A strong and vibrant community
- 13,594 developers from at least 1,340 companies have contributed to the Linux kernel since 2005
- >1,600 developers contributed to each kernel version
- A new major kernel release occurs every 9 – 10 weeks
- The Linux kernel community merged changes at an average rate of 7.8 patches per hour over the past 15 months.

A thriving commercial ecosystem
- 95%+ of the top 1M web domains
- 80%+ of smartphones run Android (based on the Linux kernel)
- 98%+ of the top 500 fastest supercomputers in the world
- Most of the global markets, including NYSE, NASDAQ, London Exchange, Tokyo Stock Exchange
- The majority of consumer electronic devices
- 75%+ of cloud-enabled enterprises report using Linux as their primary cloud platform
- E-commerce giants Amazon, EBay, PayPal, Walmart, and more run on Linux"[1]

**Legal**: The Linux kernel is released under GNU General Public License (GPL) v2 which allows distribution and sale of possibly modified and unmodified versions but requires that all those copies be released under the same license and be accompanied by the complete corresponding source code.[2]

**Finance**: The Linux Foundation is a membership organization based on different levels of membership.[3] Individual memberships are $49 per year.[4] Corporate annual membership has Platinum ($500,000), Gold ($100,000) and Silver ($5,000-$20,000) levels. Silver dues are based on total corporate consolidated headcount of the parent company.[5] The Linux Foundation has over 1,000 corporate members and thousands of individual supporters.[6]

**Marketing**: As of March 2016, top corporate contributors to Linux kernel include Intel, Red Hat, IBM, Motorola, Linaro, Google, Mellanox, SUSE, AMD, Renesas Electronics, Samsung, Rockchip, Oracle, ARM, Canonical, Broadcom.[7] The Linux system is available as a product from Red Hat. Red Hat provides storage, operating system platforms, middleware, applications, management products, and support, training, and consulting services.

**Dependency hygiene**: The Linux kernel has a few dependencies listed at this website.[8]

**Use Case: OHDSI**

https://www.ohdsi.org/data-standardization/the-common-data-model/

**Author of Use Case**: GQ Zhang (UTHSC) and Richard Boyce (U Pitt)

**Short History of the Tool:** Observational Health Data Sciences and Informatics (OHDSI) and was initiated in 2013 as a follow-up to the Observational Medical Outcomes Partnership (OMOP). OHDSI is a multi-stakeholder interdisciplinary collaborative that is striving to bring out the value of observational health data through large-scale analytics. The main objective of OHDSI is to establish a research community and open source tool set for observational health data sciences that enables active engagement across multiple disciplines spanning multiple stakeholder groups.

**Governance:** OHDSI has established an international network of researchers and observational health databases with a central coordinating center housed at Columbia University. OHDSI is not a single open source software. Rather, it is a collection of open-source tools to make analysis of observational health data more rapid, rigorous, and reproducible. The OMOP common data model related standard vocabulary are the foundational tools for all OHDSI initiatives:  https://github.com/OHDSI/CommonDataModel

**Documentation:** http://www.ohdsi.org/web/wiki/doku.php?id=documentation:overview
Documentation available about how to get started with OHDSI, Common Data Model (CDM), ETL creation best practices, and Tool Specific Documentation.

**Code Quality:** Varies since a large collection of tools are developed around the OMOP CDM. https://github.com/OHDSI

**Support:** The OHDSI community provides two methods for receiving support. The community-based Discourse forum (forums.ohdsi.org) provides support for implementing OHDSI tools, proposing or participating in network research studies, or requesting information on OHDSI related topics. Technical questions specific to software are managed through issue tickets that anyone can open on the various GitHub project sites (https://github.com/ohdsi/).

**Ecosystem Collaboration:** The OHDSI consortium is organized by Projects and Workgroups, Annual Symposium and satellite events, and Community Forum http://forums.ohdsi.org.

**Security:** The technical leadership of the OHDSI collaborative holds regular discussions related to toolset security. Apache SHIRO is used for the securing the WebAPI component (https://www.ohdsi.org/web/wiki/doku.php?id=development:security).

**Legal:** Uses Open Source licenses such as Apache License 2.

**Finance:** OHDSI is an open collaborative that, as an organization, is not specifically funded by any single organization. However, numerous individuals and teams within the collaborative have received funds to focus on specific needs and topics. Funding sources have included private and public sources.

**Marketing:** The collaborative is not a business entity and so does not actively engage in marketing. There are regular discussions about disseminating knowledge and tools developed by the collaborative. There is also a specific workgroup focusing on dissemination (https://www.ohdsi.org/web/wiki/doku.php?id=projects:workgroups:dissemination-wg). Besides these mechanisms, researchers within the collaborative tend to provide acknowledgment to OHDSI in research publications and presentations (https://www.ohdsi.org/resources/publications/).

**Dependency Hygiene:** Depends on individual tools. The OHDSI WebAPI uses Maven for dependency management. The Methods library and several other projects, including patient level prediction, use R and R related toolsets such as RStudio.

**Use Case: R**

(The R Project for Statistical Computing) https://www.r-project.org/

**Author of Use Case:** Davis M (U Pitt)

**Short History of Tool:** R is a language and environment for statistical computing and graphics. It is a GNU project similar to the S language and environment which was developed at Bell Laboratories (formerly AT&T, now Lucent Technologies) by John Chambers and colleagues. R can be considered as a different implementation of S. There are some important differences, but much code written for S runs unaltered under R. R provides a wide variety of statistical (linear and nonlinear modeling, classical statistical tests, time-series analysis, classification, clustering …) and graphical techniques, and is highly extensible. The S language is often the vehicle of choice for research in statistical methodology, and R provides an Open Source route to participation in that activity. One of R's strengths is the ease with which well-designed publication-quality plots can be produced, including mathematical symbols and formulae where needed. R is available as Free Software under the terms of the Free Software Foundation's GNU General Public License in source code form. It compiles and runs on a wide variety of UNIX platforms and similar systems (including FreeBSD and Linux), Windows and MacOS.[1]

**Governance:** The current R is the result of a collaborative effort with contributions from all over the world. Since mid-1997 there has been an R Development Core Team responsible for overseeing its development, along with the R Foundation which is a not-for-profit organization working in the public interest. The R Foundation is seated in Vienna, Austria and currently hosted by the Vienna University of Economics and Business. It is a registered association under Austrian law and active worldwide. The Foundation continues to support the development of R and provides exploration of new methodologies, teaching and training of statistical computing and the organization of meetings and conferences with a statistical computing orientation. The ordinary members, of the Foundation, are elected by a majority vote of the general assembly. New ordinary members are selected based on their non-monetary contributions (code, effort …) to the R project.

**Documentation:** Various fully developed documentation is available for both statistical end-user, developer and administrators. R provides versions for the most recent released R version (R-release), a very current version for the patched release version (R-patched) and finally a version for the forthcoming R version that is still in development (R-devel). This documentation adequately covers a wide range of topics including guides for installation, writing R Extensions covers how to create your own packages, tutorials, a guide to the internal structures of R and coding standards for the core team working on R itself.
https://cran.r-project.org/manuals.html

**Code quality:** The R development core team (R Core) using the Apache Subversion as its software versioning and revision control system to maintain current and historical versions of files such as source code, web pages, and documentation. Reasonable software development and testing methodologies are employed by R Core in order to maximize the accuracy, reliability, and consistency of R's performance. While some aspects of R's development are handled collaboratively, others are handled by members of the team with specific interests and expertise in focused areas. R Core maintains guidelines for its Software Development Life Cycle (SDLC) https://www.r-project.org/doc/R-SDLC.pdf  There have also been coding standards established providing guidance, https://cran.r-project.org/doc/manuals/R-ints.html

**Support:** R provides various support options from online self-help tools to primary FAQ listings which are periodically updated to reflect very commonly asked questions by R users. The R Project also maintains a number of subscription-based email lists for posing and answering questions about R, including the general R-help email list, the R-devel list for R code development, and R- package-devel list for developers of Comprehensive R Archive Network (CRAN) packages.

**Ecosystem collaboration:** The R Foundation actively supports two conference series, organized regularly by members from the R community:
o   useR!, providing a forum to the R user community.
o   DSC, a platform for developers of statistical software.

R Core does meet, collectively and/or in smaller groups, with a level of frequency dictated by multiple factors, including taking advantage of regularly scheduled conferences where members of R Core may already be in attendance. Such conferences include those that are specific to statistical computing and R itself (http://www.r-project.org/conferences.html). These routine communications and meetings ensure that the collaborative efforts are appropriately coordinated and prioritized as ongoing development takes place.

**Security:** The R project relies heavily on the development of functional software packages by community developers, which are submitted to a central repository, CRAN. Hence, there is an inherent security risk in malicious code making their way unnoticed into the central repositories. There are a few older published papers on the potential security risk of R and general guidelines on how to handle security issues.[2] The CRAN does have some general guidelines for submitting new packages but doesn't address any security regulations.

**Legal:** R is available as Free Software under the terms of the Free Software Foundation's GNU General Public License in source code form. It compiles and runs on a wide variety of UNIX platforms and similar systems (including FreeBSD and Linux), Windows and MacOS.

**Finance:** R is supported largely by supporting members and "one-off donations" to the R Foundation. Any person or legal entity may become a supporting member of the R foundation by paying membership fees. Annual membership fees for supporting members are:
o   Supporting natural persons: EUR 25
o   Supporting institutions: EUR 250
o   Supporting benefactors: EUR 500

**Marketing:** No direct information on marketing strategies could be located. However, R has gained market share by an "evangelist" approach amongst statisticians, data analysts and others from the biomedical community. R is comparable to popular commercial statistical packages, such as SAS, SPSS, and Stata, but R is available to users at no charge under a free software license. A number of commercial software vendors are providing commercial support and/or

40 | P a g e

extensions to their products to support R extensions to their customers.

**Dependency Hygiene:** The R package itself is self-contained in general. Its "dependencies" rely on the contributions from its community of developer libraries within the central CRAN and its distribution under the terms of the Free Software Foundation's GNU. Each package dependencies are clearly document within the package summary documentation within the CRAN site.

## Use Case: REDCap

https://www.project-redcap.org
https://projectredcap.org/resources/community/
https://projectredcap.org/software/

**Author of Use Case:** GQ Zhang

**Short History of the Tool:** REDCap (Research Electronic Data Capture) is a web-based electronic data capturing (EDC) software solution and workflow methodology, enabling data capture for clinical and translational research. REDCap uses institutional hosting as a "platform" to reach out and bring value to its intended end users.

REDCap was created in 2004 at Vanderbilt University. It was originally developed to support clinical researchers who needed a secure data collection tool that met HIPAA compliance standards. The REDCap consortium was officially launched in 2006. The consortium consists of non-profit organizations interested in expanding REDCap's functionality through collaborative software development. Each partner site was given access to the codebase so that they could install their own REDCap system and offer it to their researchers. The REDCap annual in-person conferences were established in 2009, offering educational and networking opportunities to REDCap administrators around the world.

**Governance:** https://github.com/vanderbilt-redcap REDCap is not open-source software in the usual sense, as its code base is not open to an individual developer. REDCap is available at no charge to institutional partners and is restricted in use, permitted only for non-commercial research purposes. REDCap is also restricted in redistribution because Vanderbilt is the only entity that can distribute it. Any and all derived works – such as innovations or programmatic features added on by the user – are owned by Vanderbilt. [Wikipedia]

**Documentation:** Detailed documentation is available for set up and usage, but not on contributing source code. https://projectredcap.org/resources/library/
https://projectredcap.org/about/
https://projectredcap.org/resources/community/

**Code Quality:** REDCap source code is not open to the community. However, REDCap is used by 3372 Institutions in 130 Countries with 677k Projects and 926k Users. 6408 articles citing the use of REDCap.

**Support:** Contact redcap@vanderbilt.edu  https://projectredcap.org/partners/join/

**Ecosystem Collaboration:** The REDCap Consortium has two tiers of users with distinct characteristics: institutional users and investigator users. Institutional users are responsible for installing, maintaining, and providing day-to-day support of the specific instance of the REDCap for their institution.

Investigator users - the ultimate end user, work within an instance of REDCap installed at an institution to create REDCap projects for their specific data capturing needs. Data capturing workflows are configurable according to specific project needs.

REDCap can be installed in a variety of environments for compliance with such standards as HIPAA, 21 CFR Part 11, FISMA (low, moderate, high), and international standards.

**Security:** The software application employs various methods to protect against malicious users who may attempt to identify and exploit any security vulnerabilities in the system. Input data or build an online survey or database from anywhere in the world over a secure web connection with authentication and data logging.

**Legal:** REDCap Non-Profit End-User License Agreement is made by and between Vanderbilt University ("Vanderbilt"), a not-for-profit corporation duly organized and existing under the laws of Tennessee and having offices at 1207 17th Avenue South, Suite 105, Nashville, Tennessee 37212.

**Finance:** Early support of REDCap was provided by NCRR.

**Dependency Hygiene** [Deployment environment requirement]**:** Web server with PHP (PHP 5.3.0+, including support for PHP 7). Apache (any OS) or Microsoft IIS (Windows). MySQL database server (MySQL 5.0+, MariaDB 5.1+, or Percona Server 5.1+). a MySQL client (e.g., phpMyAdmin, MySQL Workbench) is required for performing installation/upgrades SMTP email server.